\documentclass[]{pasj01}

\usepackage{lscape}

\draft

\begin{document} 
	\Received{}
	\Accepted{}

\SetRunningHead{Kaneko~et~al.}{Properties of Molecular Gas in Galaxies in Early and Mid Stages of Interaction. III.}

\title{Properties of Molecular Gas in Galaxies in Early and Mid Stages of Interaction. III. Resolved Kennicutt--Schmidt Law}

\author{Hiroyuki \textsc{Kaneko},\altaffilmark{1,2}
	Nario \textsc{Kuno},\altaffilmark{3,4}
	Daisuke \textsc{Iono},\altaffilmark{2,5}
	Yoichi \textsc{Tamura},\altaffilmark{6}
	Tomoka \textsc{Tosaki},\altaffilmark{7}
	Kouichiro \textsc{Nakanishi},\altaffilmark{2,5}
	and Tsuyoshi \textsc{Sawada}\altaffilmark{2,8}
}
\email{hkaneko@juen.ac.jp}
\altaffiltext{1}{Graduate School of Education, Joetsu University of Education, 1 Yamayashiki-machi, Joetsu, Niigata, 943-8512, Japan}
\altaffiltext{2}{National Astronomical Observatory of Japan, 2-21-1 Osawa, Mitaka, Tokyo, 181-8588, Japan}
\altaffiltext{3}{Graduate School of Pure and Applied Sciences, University of Tsukuba, 1-1-1 Tennodai, Tsukuba, Ibaraki, 305-8577, Japan}
\altaffiltext{4}{Tomonaga Center for the History of the Universe, University of Tsukuba, Tsukuba, Ibaraki 305-8571, Japan}
\altaffiltext{5}{Department of Astronomical Science, The Graduate University for Advanced Studies, 2-21-1 Osawa, Mitaka, Tokyo, 181-8588, Japan}
\altaffiltext{6}{Division of Particle and Astrophysical Science, Graduate School of Science, Nagoya University, Furo-cho, Chikusa-ku, Nagoya, Aichi, 464-8602, Japan}
\altaffiltext{7}{Department of Geoscience, Joetsu University of Education, 1, Yamayashiki-machi, Joetsu, Niigata, 943-8512, Japan}
\altaffiltext{1}{Graduate School of Pure and Applied Sciences, University of Tsukuba, 1-1-1 Tennodai, Tsukuba, Ibaraki, 305-8577, Japan}\altaffiltext{8}{Joint ALMA Observatory, Alonso de Cordova 3107, Vitacura, Santiago, Chile}


%


\maketitle

\begin{abstract}	
We study properties of the interstellar medium, an ingredient of stars, and star formation activity, in four nearby galaxy pairs in the early and mid stages of interaction for both a galaxy scale and a kpc scale. 
The galaxy-scale Kennicutt--Schmidt law shows that seven of eight interacting galaxies have a star formation rate within a factor of three compared with the best-fit of the isolated galaxies, although we have shown that molecular hydrogen gas is efficiently produced from atomic hydrogen during the interaction in the previous paper.
The galaxy-scale specific star formation rate (sSFR) and star formation efficiency (SFE) in interacting galaxies are comparable to those in isolated galaxies.
We also investigate SFE and the Kennicutt--Schmidt law on a kpc scale.
The spatial distributions of SFE reveal that SFE is locally enhanced, and the enhanced regions take place asymmetrically or at off-centre regions.
The local enhancement of SFE could be induced by shock.
We find that the index of the Kennicutt--Schmidt law for the interacting galaxies in the early stage is 1.30$\pm$0.04, which is consistent with that of the isolated galaxies.
Since CO emission, which is used in the Kennicutt--Schmidt law, is a tracer of the amount of molecular gas, this fact suggests that dense gas, which is more directly connected to star formation, is not changed at the early stage of interaction.
\end{abstract}

\section{Introduction}
Galaxy-galaxy interactions are one of the fundamental phenomena for galaxy evolution, as proved by the increase of merger rate along with the red-shift (e.g., \cite{BCS10}).
They significantly alter the morphological, dynamical, and chemical features of progenitors.
Compared with isolated spiral galaxies, galaxies under close interaction with other galaxies tend to have higher star formation activity traced by H$\alpha$ \citep{Bushouse87}, radio continuum \citep{Stocke78}, and infrared (IR: \cite{BLW88}) emission.
What makes these drastic changes is an important issue for understanding the effects of the galaxy-galaxy interaction.

In this context, the properties of molecular gas are key to understanding the star formation in galaxies because cold molecular clouds fuel current and future star formation.  
Most of (ultra-)luminous infrared galaxies (U/LIRGs, whose IR luminosities of $L_{\rm IR} > 10^{12} \LO$ and $> 10^{11} \LO$, respectively) found in the local Universe are interacting galaxies \citep{Sanders88}.
Therefore, a cause of active star formation in interacting galaxies have been examined with the relationship between star formation rate (SFR) and molecular gas for U/LIRGs.
\citet{Young96}, one of the pioneering works for this realm, exhibited that interacting galaxies have higher star formation efficiency (SFE: a ratio of SFR and molecular gas mass) than isolated galaxies. 
\citet{Violino18} also statistically confirmed that close interacting galaxies have shorter depletion time (an inverse of SFE).
From observations of nearby LIRGs with dense molecular gas tracers such as high--$J$ CO, HCN, and HCO$^{+}$ lines,
it is suggested that a higher fraction of dense gas in interacting galaxies is likely to be the cause of their high SFE \citep{GS04,Liu15,Michiyama16}.
 
Another key investigation for the relationship between SFR and cold interstellar gas is the Kennicutt--Schmidt law \citep{Schmidt59,Kennicutt98a}.
Observational studies on a galaxy scale have shown that the surface density of SFR and that of the interstellar medium (ISM) obey a power-law relation:
\begin{equation}
	{\rm log}\ \Sigma_\mathrm{SFR}= A + N{\rm log}\ \Sigma_\mathrm{ISM}.
	\label{SKlaw}
\end{equation}
Observational studies of nearby isolated galaxies show $N$ of 1.2--1.4 in both a galaxy and a kpc scale.
\citet{Daddi10} derived the Kennicutt--Schmidt law for local and high-z galaxies.
They found that starburst galaxies like U/LIRGs form stars rapidly, while spiral galaxies and BzK galaxies show a long-lasting star formation.
This result implies the physical mechanism of star formation in interacting galaxies is different from isolated spiral galaxies.
  
However, in most cases, these observational studies discussed only the galaxy-scale physical conditions due to the limitation of spatial resolution.
Although there are observations that can resolve galactic structures, most of them focused on the interacting galaxies that already show active star formation.
This bias hinders pursuing where and when star formation is enhanced during the interaction.
Many numerical simulations have illustrated that molecular gas inflow is induced by galaxy-galaxy interactions \citep{BH96,TCB10,Hopkins13}.
The inflow makes the density of molecular gas high at the central region of galaxies, and then such molecular gas efficiently converts into stars.
Observational studies reported that the inflow occurs in some interacting galaxies (e.g., \cite{Iono04}).
However, not all the interacting galaxies with active star formation show the central starburst.
For example, the most active star-forming site of the Antennae Galaxies, one of the representative interacting galaxies in the local Universe, are not at their centres but their overlap region \citep{Whitmore99}.
For understanding the effect of a galaxy-interaction event on an enhancement of star formation, three aspects should be important: 
(1) before or at the onset of the enhancement of star formation (i.e., not being a merging phase), (2) molecular gas as an ingredient for star formation, and (3) resolving galactic structures.

Following the strategies listed above, \citet{Kaneko13} (Paper I) observed four interacting galaxies in the early and mid stages of the interaction in \atom{CO}{}{12}(\textit{J}~=~1--0) (hereafter, we denote CO) emission line.
Since they mapped in kpc scale, which can resolve galactic structures, it is suitable for investigations of properties of the ISM under the effect of the interaction.
Using Paper I data, \citet{Kaneko17} (hereafter Paper II) illustrated that a galaxy-scale molecular gas fraction in interacting galaxies is higher than isolated galaxies, even in the early stage of the interaction.
Furthermore, they showed that high molecular gas fraction is due to an efficient conversion from atomic gas to molecular gas by external pressure using theoretical model fitting.
Since molecular gas is fuel for stars, these results imply that star formation activity in these interacting galaxies is also affected by the interactions.

In this paper, we focus on relationships between gas contents and star formation activity.
The structure of this paper is as follows: We explain the data we used, including isolated spiral galaxies for comparison in section 2.
Section 3 presents derived values and describe the derivation of the dust-extinction corrected star formation rate. 
We discuss star formation activity such as specific star formation rate, star formation efficiency, and the Kennicutt--Schmidt law on a galaxy scale in section 4. 
Then we discuss the kpc-scale SFR and SFE in section 5. 
The spatially-resolved Kennicutt--Schmidt law is discussed in section 6.
Finally, we summarise this study in section 7.

\section{Data}\label{data}
We collect data covering an entire system of interacting galaxies in order to understand a relationship between ISM contents and star formation properties in interacting galaxies.
We set a condition on the sample to resolve galactic structures (a linear resolution of $<$ 10 kpc).
We select sample galaxy pairs that are in the early and the mid stages of the interaction.
The stage definition is based on \citet{Zhu03}.
The early stage of the interaction is defined by the separation of nuclei and morphology.
The interacting galaxies in the early stage fulfil $R < 1.5D_{25}$, where $R$ is the projected separation between the nuclei, and $D_{25}$ is the diameter of the primary galaxy, using the brightness contour, $B_\mathrm{T}$ = 25 mag arcsec$^{-2}$.
In addition, they also show small morphological disturbances. 
Therefore, we classify galaxies contacting the celestial sphere and having a severely disturbed morphology, i.e., colliding galaxies, as the mid stage of interaction.
We obtain data of molecular gas (CO), atomic hydrogen gas (H\emissiontype{I}), old stars ($K_\mathrm{s}$-band), and star formation tracers (H$\alpha$, FUV, 8 $\mu$m and 24 $\mu$m).
We also use a dataset of CO, H\emissiontype{I}, $K_\mathrm{s}$-band, H$\alpha$, FUV, 8 $\mu$m, and 24 $\mu$m images of isolated spiral galaxies for comparison.

\subsection{CO data}
\begin{figure*}[tbp]
	\centering
	\includegraphics{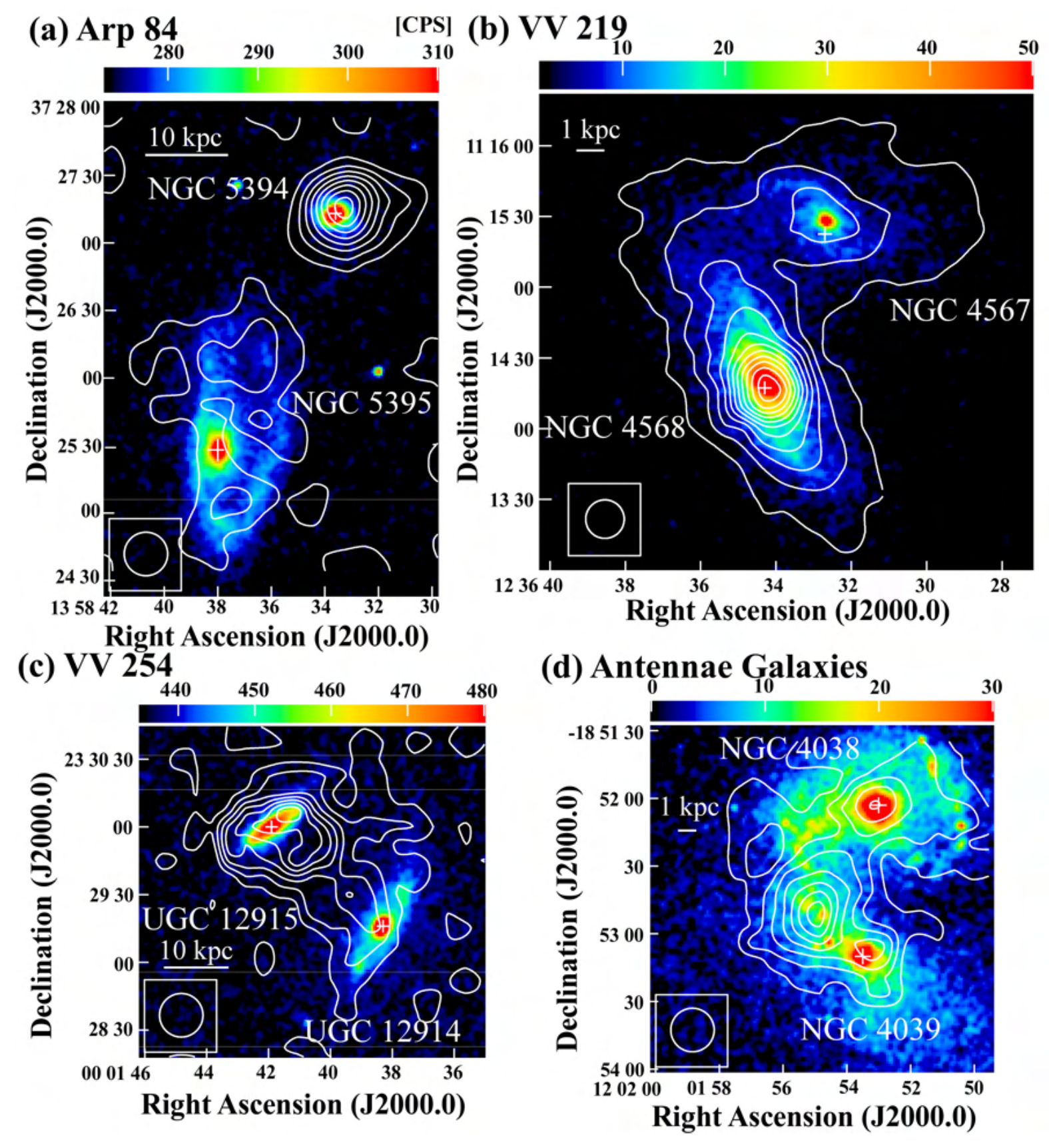}
	\caption{2MASS $K_\mathrm{s}$-band image with CO integrated intensity contours for interacting galaxies. Circles in the bottom-left corner of each panel are the beam size of CO (\timeform{19''.3}). 
		(a) Arp\,84 image. The CO contour levels are 2.81 $\times$ 2, 3, 4, ... K km s$^{-1}$. The crosses represent the galactic centres of NGC\,5394 (top right) and NGC\,5395 (bottom left).
		(b) The same as (a) but for VV\,219. The contours are 7.20 $\times$ 1, 2, 3, ... K km s$^{-1}$. The crosses represent the galactic centres of NGC\,4567 (top right) and NGC\,4568 (bottom left).
		(c) The same as (a) but for VV\,254. The contours are 6.63 $\times$ 1, 2, 3, 4, ... K km s$^{-1}$. The crosses represent the galactic centres of UGC\,12914 (bottom right) and UGC\,12915 (top left).
		(d) The same as (a) but for the Antennae Galaxies. The contours are 23.7 $\times$ 1, 2, 3, ... K km s$^{-1}$. The crosses represent the galactic centres of NGC\,4038 (top) and NGC\,4039 (bottom).}
	\label{Ks}
\end{figure*}

We used CO data for interacting galaxies from Paper I.
The dataset was obtained with the Nobeyama 45-m radio telescope whose effective angular resolution is \timeform{19''.3} and pixel size is 7$\farcs$5.
An interacting galaxy sample consists of four systems: Arp\,84, VV\,219, VV\,254, and the Antennae Galaxies.
The sensitivities of the data are $3\times 10^{-3}$--$1\times 10^{-2}$ Jy km s$^{-1}$ arcsec$^{-2}$.
Based on our definition, Arp\,84, VV\,219, and VV\,254 are the early stage of interaction, while the Antennae Galaxies are in the mid stage.
The similar recession velocities of these galaxy pairs suggest they are clearly colliding systems as opposed to apparent close pairs due to projection effects.
Figure \ref{Ks} shows $K_\mathrm{s}$-band images with the CO contours of interacting galaxies.
Details of the CO data are summarised in table \ref{interacting}.

We used the CO data for isolated spiral galaxies from Nobeyama CO Atlas \citep{Kuno07} for comparison.
Since the CO dataset of this atlas was obtained with the Nobeyama 45-m radio telescope as our interacting galaxy sample, it is a suitable CO dataset for comparison purposes. 
We only used galaxies that all other tracers are available.
To remove other environmental effects, we excluded the galaxies belonging to the Virgo and Coma clusters.
As a result, 11 of 40 galaxies were used for analysis (see table \ref{isolate}).

Both datasets are obtained with the on-the-fly (OTF) method, resulting in almost constant noise throughout the map.
The typical systematic error of intensity calibration for the Nobeyama 45-m telescope is about 15\%. 
The data with the pointing error worse than 5$^{\prime\prime}$, which is higher than the grid spacing of 7$\farcs$5, are removed in the data reduction.

\begin{table*}[tbp]
	\tbl{Interacting galaxies.}{%
		\begin{tabular}{cccccccc}
			\hline
			Pair Name & 	Galaxy			   &   Morphology &  Distance  & Inclination & Orbit & $R_\mathrm{K20}$ & Resolution\\
				      &                        &              &  (Mpc)     &   (deg)     &       & (arcsec) & (kpc)\\
			(1)       &       (2)              &   (3)        &    (4)     &     (5)     &  (6)  &  (7)  & (8)\\
			\hline
			Arp\,84 	  		  & NGC\,5394  &  SB(s)b pec      &  53  &   0  &   P   &    34    & 5.0\\
					 		  &	NGC\,5395  &  SA(s)b pec      &      &  58  &   R   &    71    & \\
			\hline 
			VV\,219	  		  &	NGC\,4567  &   SA(rs)bc       &  16  &  44  &   R   &    70    & 1.5\\
					  		  &	NGC\,4568  &   SA(rs)bc       &      &  58  &   R   &    94    & \\
			\hline 
			VV\,254	  		  &	UGC\,12914 &  (R)S(r)cd pec   &  62  &  61  &   H   &    55    & 5.8\\
					  		  &	UGC\,12915 &     S?           &      &  73  &   H   &    40    & \\
			\hline
			Antennae Galaxies & NGC\,4038  &  SB(s)m pec      &  21  &  --- &   P   &    73    & 2.0\\
					 		  &	NGC\,4039  &  SA(s)m pec      &      &  --- &   P   &    81    & \\
			\hline
		\end{tabular}}
		\label{interacting}
		\begin{tabnote}
			Note --- 
			Column (1): Pair name.
			Column (2): Name of the constituent galaxy.
			Column (3): Morphological type from NED.
			Column (4): Distance to the pair.
			Column (5): Inclination angle. Reference: \citet{Kaufman99} for Arp\,84, this work for VV\,219 and \citet{Giovanelli86} for VV\,254. Since the Antennae Galaxies have complex morphologies, the inclination is not fixed.
			Column (6): Orbital type judging from their morphology; P, R, H represent a prograde, retrograde and head-on collision, respectively.
			Column (7): The fitted radius at 20 mag arcsec$^{-2}$ in the $K_\mathrm{s}$-band in Paper I.
			Column (8): Linear spatial resolution corresponding to the effective angular resolution of CO, \timeform{19''.3}, at the distance of galaxy pairs. 
	\end{tabnote}
\end{table*}

\begin{table*}[tbp]
	\tbl{Control isolated galaxies.}{%
	\begin{tabular}{cccccc}
		\hline
		Name    &   Morphology    &   Velocity    &  Distance   & Inclination  	& $R_\mathrm{K20}$\\
		&                 & (km s$^{-1}$) &    (Mpc)    &    (deg)  	& (arcsec)   \\
		(1)     &      (2)        &      (3)      &     (4)     &     (5)		& (6)\\
		\hline
		NGC\,253		& SAB(s)c		&    227	&     2.5	&      75	&	630.2\\
		NGC\,2903	& SAB(rs)bc		&    549	&     6.3	&      67	&	163.0\\
		NGC\,3184	& SAB(rs)cd		&    594	&     8.7	&      21	&	114.6\\
		NGC\,3351	& SB(r)b		&    778	&    10.1	&      40	&	116.3\\
		NGC\,3521	& SAB(rs)bc		&    792	&     7.2	&      63	&	164.4\\
		NGC\,3627	& SAB(s)b		&    715	&    11.1	&      52	&	185.0\\
		NGC\,4736	& (R)SA(r)ab	&    317	&     4.3	&      40	&	172.3\\
		NGC\,5055	& SA(rs)bc		&    503	&     7.2	&      61	&	204.2\\
		NGC\,5236	& SAB(s)c		&    514	&     4.5	&      24	&	312.4\\
		NGC\,5457	& SAB(rs)cd		&    255	&     7.2	&      18	&	236.3\\
		NGC\,6946	& SAB(rs)cd		&     60	&     5.5	&      40	&	252.5\\
		\hline
	\end{tabular}}
	\label{isolate}
	\begin{tabnote}
		Note --- 
		Column (1): Galaxy name.
		Column (2): Morphological type.
		Column (3): Velocity in local standard of rest \citep{Kuno07}.
		Column (4): Distance.
		Column (5): Inclination angle.
		Column (6): Radius at 20 magnitude arcsec$^{-2}$ in the $K_\mathrm{s}$-band.
	\end{tabnote}
\end{table*}

\subsection{H\emissiontype{I} data}
All H\emissiontype{I} data for interacting galaxies were acquired with the Very Large Array (VLA)\footnote{The National Radio Astronomy Observatory is a facility of the National Science Foundation operated under cooperative agreement by Associated Universities, Inc.}.
The H\emissiontype{I} images for Arp\,84, VV\,219, VV\,254, and the Antennae Galaxies were taken from the VLA archive, \citet{Condon93}, \citet{Iono05}, and \citet{Hibbard01}, respectively.
Since the VLA is an interferometer, the calibration error is nearly constant throughout a map.
The angular resolutions of the H\emissiontype{I} data are typically 10$^{\prime\prime}$--15$^{\prime\prime}$, and all of them are higher than CO data.

For isolated galaxies, we used the H\emissiontype{I} data obtained with the VLA by THINGS\footnote{This work made use of THINGS, `The H\emissiontype{I} Nearby Galaxy Survey' \citep{Walter08}} except for NGC\,253, whose data was taken with the Australia Telescope Compact Array \citep{Boomsma05}.
The THINGS datasets have two weighting images: natural weighting and robust weighting.
We used the natural weighting, which leads to higher sensitivity with lower angular resolution.
Each THINGS H\emissiontype{I} data of isolated galaxies has an angular resolution of about 10$^{\prime\prime}$--15$^{\prime\prime}$ with a grid spacing of 1$\farcs$5.
The main uncertainty for H\emissiontype{I} data is flux calibration error ($\sim$5\%) and continuum subtraction (depending on the data, but low-quality data is not in public: \cite{Walter08}).
Since the H\emissiontype{I} data of NGC\,253 has an angular resolution of 70$^{\prime\prime}$, which is much larger than other samples, we only use this data to discuss the galaxy scale.

\subsection{$K_\mathrm{s}$ data}
In order to derive stellar mass, we used the $K_\mathrm{s}$-band images.
All data were taken from the Two Micron All Sky Survey (2MASS) catalogue\footnote{This publication makes use of data products from the Two Micron All Sky Survey, which is a joint project of the University of Massachusetts and the Infrared Processing and Analysis Center/California Institute of Technology, funded by the National Aeronautics and Space Administration and the National Science Foundation} \citep{Skrutskie06, Jarrett03}.
The point spread function (PSF) for $K_\mathrm{s}$ is about 2$^{\prime\prime}$--3$^{\prime\prime}$. 
Photometric zero-point calibration is accurate to 2-3\%, and the relative intensity calibration error of 2MASS data is uniformly 2-3\% over the sky.

\subsection{H$\alpha$ and FUV data}
We collected H$\alpha$ data from published papers \citep{Koopmann01,Xu00}.
{The typical systematic flux calibration errors, including continuum subtraction and calibration errors, are estimated to be 20-30\% in total H$\alpha$ fluxes.
Since Arp\,84 has no available H$\alpha$ image, we used the Galaxy Evolution Explorer (GALEX) FUV images retrieved from MultiMission Archive at Space Telescope Science Institute (MAST).
The original observations were a part of the GALEX Nearby Galaxies Survey (NGS) \citep{Gil_de_Paz07}.

For isolated galaxies, H$\alpha$ images are obtained from \citet{Kennicutt08} except for NGC\,3184, which was observed by \citet{Young96}. 
	The data reduction processes are similar to \citet{Koopmann01}.

The PSFs of H$\alpha$ and FUV for interacting galaxies and isolated galaxies are typically 2$^{\prime\prime}$ and 4$^{\prime\prime}$--4$\farcs$5, respectively.
These are much smaller than the angular resolution of CO and H\emissiontype{I} data.

\subsection{MIPS 24 $\mu$m and IRAC 8 $\mu$m data}
Star formation activity estimated from only H$\alpha$ or FUV is underestimated due to dust absorption.
For correcting the dust absorption, we used the 24 $\mu$m Multiband Imaging Photometer (MIPS: \cite{Rieke04}) and 8 $\mu$m Infrared Array Camera (IRAC: \cite{Fazio04})
datasets of the basic calibrated data (BCD) created by the Spitzer Science Center (SSC) pipeline from the Spitzer Space Telescope \citep{Werner04} data archive\footnote{This work is based on observations made with the Spitzer Space Telescope, which is operated by the Jet Propulsion Laboratory, California Institute of Technology under a contract with NASA.}.
For IRAC 8 $\mu$m and MIPS 24 $\mu$m datasets, thermal noise and calibration uncertainty are typically less than 10\% (also see \cite{Smith07}).
The PSFs of IRAC 8$\mu$m and MIPS 24 $\mu$m for all samples are about 2$^{\prime\prime}$ and 5$^{\prime\prime}$, respectively.

\section{Derived properties of molecular gas, atomic gas, stars, and star formation activity}\label{values}
We derive molecular hydrogen, atomic hydrogen, stellar mass, and SFR from collected data.
Molecular hydrogen gas mass is derived from CO data, assuming the Galactic $I_\mathrm{CO}-N_\mathrm{H_{2}}$ conversion factor of 1.8 $\times$
10$^{20}$ [cm$^{-2}$ (K km s$^{-1}$)$^{-1}$] \citep{Dame01}.
The uncertainties in the integrated intensity were estimated to be less than 15\% for each map due to the rms error in the spectrum and the error in the determination of the baseline (the details are described in Paper I).

Since the conversion factor varies from one galaxy to another and within each galaxy \citep{Sandstrom13}, the scatter and the uncertainty of the Kennicutt--Schmidt relation become larger when a unique conversion factor is used.
From both observational and theoretical approaches, the conversion factor gets lower in high surface density environments such as ULIRGs \citep{Bolatto13}.
Although our sample interacting galaxies are not CO-bright compared with ULIRGs, some targets are known to have lower conversion factors \citep{Zhu03,Zhu07}. 
If we suppose that the conversion factor in all our sample interacting galaxies is lower than the Galactic conversion factor, then the measurements of the surface density of molecular gas and total gas are overestimated.
The lower conversion factor makes SFE and the slope of the Kennicutt--Schmidt law higher than the values we will investigate.

We do not take into account the errors, including the observational uncertainties and the variation of the $I_\mathrm{CO}-N_\mathrm{H_{2}}$ conversion factor in deriving the index and coefficient of the Kennicutt--Schmidt law.
In spiral galaxies, the standard deviation of the $I_\mathrm{CO}-N_\mathrm{H_{2}}$ conversion factor is 0.3 dex \citep{Sandstrom13}.
So, if this variation also suits to interacting galaxies, we expect the surface density of molecular gas has an uncertainty factor of at least 0.4 dex.

Mass of atomic hydrogen gas is calculated from H\emissiontype{I} assuming an optically thin emission, $M_\mathrm{H\emissiontype{I}}[\MO] = 2.36 \times 10^{5} D^{2} S_\mathrm{H\emissiontype{I}}$, where $D$ is the distance in megaparsec, and $S_\mathrm{H\emissiontype{I}}$ is the integrated flux of H\emissiontype{I} in Jy km s$^{-1}$.
Stellar mass is derived from $K_\mathrm{s}$-band luminosity using a mass-to-light ratio of 0.95 ($M_{*}$/$L_{K_\mathrm{s}}$) [\MO/\LO] \citep{Bell03}.
The derived fundamental quantities from the collected data for interacting galaxies and isolated galaxies are summarised in tables \ref{IG} and \ref{COatlas}, respectively.

\begin{table*}[tbp]
	\tbl{Basic properties of interacting galaxies.}{%
	\begin{tabular}{ccccccc}
		\hline
		Galaxy  &  $M_\mathrm{H_{2}}$  & $M_\mathrm{H\emissiontype{I}}$ &	$M_{*}$ & $L_\mathrm{H\alpha}$($L_\mathrm{FUV}$) & $L_{24\mu m}$($L_{8\mu m}$) & Reference\footnotemark[$\dagger$] \\
				& (10$^{9} \ \MO$) &   (10$^{9} \ \MO$) &	(10$^{9} \ \MO$) &	(10$^{40(42)}$ erg s$^{-1}$) &   (10$^{42}$ erg s$^{-1}$) &\\
   		(1)     & (2)           & (3)            & (4)           &  (5)       & (6)    &(7) \\
		\hline 
		NGC\,5394   	& 4.8		&	0.08	&	7.9		&	7.4 (5.1)		& 2.9		& 1, 3, 6 \\
		NGC\,5395   	& 8.0		&	5.7		&	26.7	&	1.3 (3.1)		& 10.7		& 1, 3, 6 \\
		NGC\,4567   	& 1.3		&	0.2		&	1.7		&	3.8				& 0.8  		& 1, 4, 6 \\
		NGC\,4568   	& 3.0		&	1.1		&	5.2		&	7.9				& 2.0  		& 1, 4, 6 \\
		UGC\,12914	& 10.3		&	6.2		&	20.5	&	5.1				& 8.3  		& 1, 3, 7 \\
		UGC\,12915	& 14.9		&	5.1		&	11.6	&	3.1				& 4.9		& 1, 3, 7 \\
		NGC\,4038  	& 6.2		&	2.5		&	8.2		&	47.2			& 2.7		& 2, 5, 7 \\
		NGC\,4039   	& 3.2		&	1.0		&	7.2		&	50.8			& 2.1		& 2, 5, 7 \\
		\hline
	\end{tabular}}
	\label{IG}
	\begin{tabnote}
		Note --- 
		Column (1): Galaxy name.
		Column (2): Mass of molecular hydrogen.
		Column (3): Mass of atomic hydrogen.
		Column (4): Stellar mass.
		Column (5): H$\alpha$ luminosity. For Arp\,84 and VV\,254, FUV luminosity.
		Column (6): 24 $\mu$m luminosity. For Arp\,84, 8 $\mu$m luminosity. $L_{24\mu m}$ and $L_{8\mu m}$ are monochromatic luminosity, $\nu L_\nu$. For the 8.0 and 24 $\mu$m bands, the effective frequencies used were 8.45 $\times$ 10$^{13}$ and 1.27 $\times$ 10$^{13}$ Hz.
		Column (7): References of the H\emissiontype{I}, H$\alpha$, FUV, 24 $\mu$m, and 8 $\mu$m data.\\
		References: 1.\citet{Iono05}; 2.\citet{Hibbard01}; 3. GALEX NGS \citep{Gil_de_Paz07};
		4.\citet{Koopmann01}; 5.\citet{Xu00}; 6. \citet{Smith07}; 7. Spitzer archive.
	\end{tabnote}
\end{table*}

\begin{table*}[tbp]
	\tbl{Basic properties of isolated galaxies.}{%
	\begin{tabular}{ccccccc}
		\hline
		Galaxy  &  $M_\mathrm{H_{2}}$  & $M_\mathrm{H\emissiontype{I}}$  &	$M_{*}$ &  $L_\mathrm{{H\alpha}}$   & $L_{24\mu m}$ & Reference \\ 
		&  (10$^{9}\MO$) & (10$^{9}\MO$) &	(10$^{9} \ \MO$) & (10$^{40}$ erg s$^{-1}$)  & (10$^{42}$ erg s$^{-1}$) &\\ 
		(1)    &          (2)           &      (3)            &         (4)           &                 (5)       &              (6)        & (7)      \\ 
		\hline
		NGC\,253  &  1.6  & 1.0  &	4.7	& 3.9  & 13.7 & 1, 4, 7 \\
		NGC\,2903 &  1.6  & 2.2  &	3.5	& 5.9  & 4.2  & 2, 4, 7 \\
		NGC\,3184 &  1.1  & 1.9  &	1.5	& 6.9  & 1.6  & 2, 3, 5 \\
		NGC\,3351 &  1.1  & 1.2  &	4.0	& 4.6  & 3.7  & 2, 4, 6 \\
		NGC\,3521 &  2.9  & 3.6  &	5.4	& 8.1  & 4.1  & 2, 4, 5 \\
		NGC\,3627 &  8.8  & 1.2  &	1.3	& 15.7 & 13.4 & 2, 4, 6 \\
		NGC\,4736 &  0.4  & 0.3  &	4.0	& 2.4  & 1.5  & 2, 4, 5 \\
		NGC\,5055 &  2.9  & 6.8  &	6.2	& 9.8  & 4.5  & 2, 4, 5 \\
		NGC\,5236 &  2.0  & 1.7  &	6.2	& 18.3 & 12.0 & 2, 4, 7 \\
		NGC\,5457 &  3.2  & 14.2 &	4.7	& 24.7 & 8.2  & 2, 4, 7 \\
		NGC\,6946 &  4.0  & 3.6  &	5.9	& 25.1 & 9.1  & 2, 4, 5 \\
		\hline
	\end{tabular}}
	\label{COatlas}
	\begin{tabnote}
		Each column is described as the same manners as table \ref{IG}.\\
		References: 1. \citet{Boomsma05}; 2. \citet{Walter08}; 3. \citet{Young96}; 4. \citet{Kennicutt08}; 5. \citet{Smith07}; 6. \citet{Dale05}; 7. \citet{Dale09}		
	\end{tabnote}
\end{table*}
		
We derived dust-extinction corrected SFR.
We used three calibration methods to correct the dust extinction;
that is, we calculated SFR using H$\alpha$ (or FUV for Arp\,84 and VV\,254) and MIPS 24 $\mu$m (or IRAC 8 $\mu$m for Arp\,84) data.
Based on the method proposed by \citet{Calzetti07}, the extinction of H$\alpha$ by surrounding dust is corrected using Spitzer MIPS 24 $\mu$m data. 
We also used similar methods for FUV emission using Spitzer MIPS 24 $\mu$m or IRAC 8 $\mu$m data \citep{Zhu08}.
The calibration equations are as follows:
\begin{eqnarray}
	L_\mathrm{H\alpha(corr)}  &=& L_\mathrm{H\alpha(obs)}+(0.032\pm0.006)L_\mathrm{24\mu m}\\
	L_\mathrm{FUV(corr)}       &=& L_\mathrm{FUV(obs)}+6.31L_\mathrm{24\mu m}\\
&=& L_\mathrm{FUV(obs)}+3.02L_\mathrm{8\mu m},
\end{eqnarray}
where the luminosities are in erg s$^{-1}$, and $L_\mathrm{FUV}$, $L_\mathrm{24 \mu m}$ and $L_\mathrm{8 \mu m}$ are monochromatic luminosity $\nu L_{\nu}$.
$\nu$ is the effective frequency and $L_{\nu}$ is the luminosity of the source at the image.
Since \citet{Calzetti07} report that the extinction correction using 24 $\mu$m is better than using 8 $\mu$m, we used 24 $\mu$m data except for Arp\,84.
For Arp\,84, we used 8 $\mu$m data to calculate SFR because 24 $\mu$m emission is saturated in NGC\,5394, one of the constituent galaxies of Arp\,84.

After the extinction correction, SFR was calculated from H$\alpha$ using the equation \citet{Calzetti07} proposed. 
To calculate SFR from FUV data, we adopt the equation introduced by \citet{BrC03} instead of \citet{Kennicutt98b} 
because it is adjusted for the GALEX FUV data.
These equations are as follows:
\begin{eqnarray}
	\label{deriveSFR1}
	\mathrm{SFR} \ [\MO\, \mathrm{yr^{-1}]} &=& 5.3 \times 10^{-42} L_\mathrm{H\alpha(corr)} \ [\mathrm{erg \ s^{-1}}] \\
	\mathrm{SFR} \ [\MO\,\ \mathrm{yr^{-1}]} &=& 6.4 \times 10^{-41} L_\mathrm{FUV(corr)} \ [\mathrm{erg \ s^{-1}}].
	\label{deriveSFR2}
\end{eqnarray}

In this paper, we do not include the systematic errors introduced by observations in the error estimation.
As described in section \ref{data}, the systematic errors for each data range from a few to 30\% due to flux calibration, continuum subtraction, or contamination of [N\emissiontype{II}] lines.
Therefore, we expect the derived SFR to have an uncertainty factor of 0.4 dex, which is the same as the molecular gas mass estimate.

\section{Galaxy-scale star formation activity}
\label{GSF}
Previous works have discussed differences in the galaxy-scale star formation activities, ISM, and stellar contents between interacting galaxies and isolated galaxies (e.g., \cite{Young96}; \cite{Saintonge11}; \cite{Pan18}).
In order to check whether our sample shows enhanced star formation activity on a galaxy scale, we compare galaxy-scale star formation activities with stellar and interstellar gas components.

\begin{figure}[tbp]
	\centering
	\includegraphics{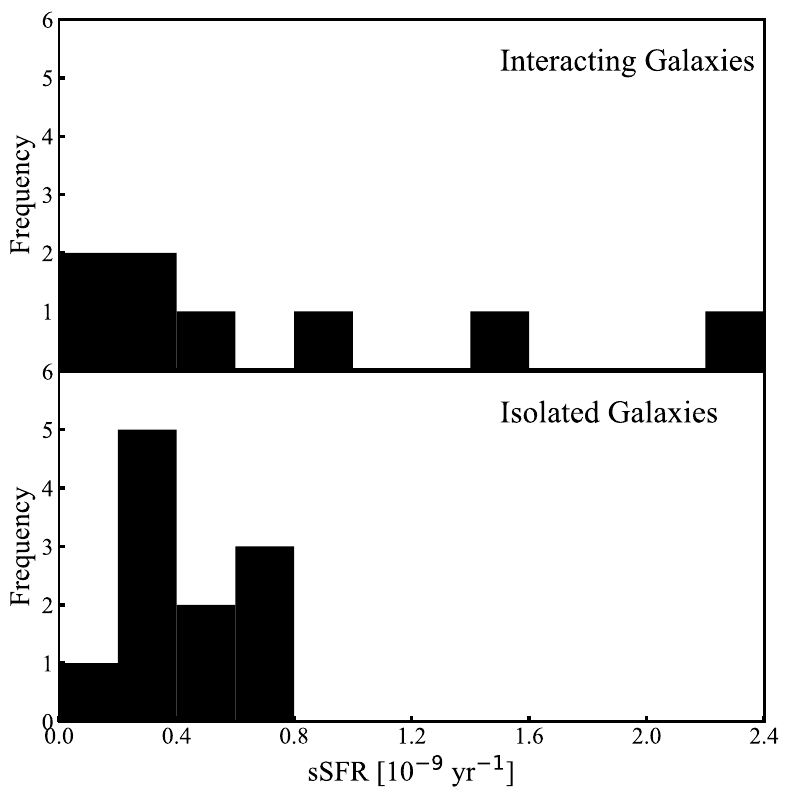}
	\caption{Histograms of specific star formation rate for the interacting galaxies (upper panel) and the isolated galaxies (lower panel).}
	\label{sSFRhist}
\end{figure}

First, we examine the specific star formation rate (sSFR), defined as SFR per stellar mass, SFR/$M_{*}$.
The sSFR represents recent star formation activity.
The histogram of sSFR is shown in figure \ref{sSFRhist}.
The dispersion of sSFR in interacting galaxies is larger than in isolated galaxies (tables \ref{derived_interact} and \ref{derived_isolated}).
NGC\,4038 and NGC\,4039, both of which are a part of the Antennae Galaxies, have sSFR higher than 1.0 $\times$ 10$^{-9}$ yr$^{-1}$.
Interacting galaxies have an average of sSFR of (7.0$\pm$5.0) $\times$ 10$^{-10}$ yr$^{-1}$,
while that of the isolated galaxies is (4.1$\pm$2.0) $\times$ 10$^{-10}$ yr$^{-1}$.
The median of sSFR is 4.0 $\times$ 10$^{-10}$ yr$^{-1}$ for interacting galaxies and 3.4 $\times$ 10$^{-10}$ yr$^{-1}$ for the isolated galaxies.
The Kolmogorov--Smirnov test for sSFR shows that the current star formation activity of interacting galaxies in the early stage, on average, is as same as isolated galaxies ($p$-value of 0.42).
The result is consistent with previous studies \citep{KCQ15,Pan18} and implies interacting galaxies in the early stage do not show the burst of star formation as we expected that they are before or at the onset of the enhancement of star formation.

Although many studies suggest that molecular gas governs star formation, the contribution of atomic gas (H\emissiontype{I}) on star formation is still under debate.
\citet{Bigiel08} found that the surface density of H\emissiontype{I} gas saturates around $\Sigma_\mathrm{ISM} \simeq10\,\MO$\,pc$^{-2}$ and it is not connected to the star formation.
Takeuchi \etal (private communication) reached the same conclusion using a larger dataset.
However, recent studies suggest atomic hydrogen directly contributes to star formation activity \citep{Fukui19}.
Therefore, this paper will investigate the relationship between star formation and two ISM masses: molecular hydrogen gas and total hydrogen gas (a sum of H\emissiontype{I} and H$_{2}$ masses).

\begin{table}[tbp]
	\tbl{Galaxy-scale star formation properties of interacting galaxies.}{%
	\begin{tabular}{ccccc}
		\hline
		Galaxy & SFR                     &    sSFR			 & SFE$_\mathrm{mol}$ & SFE$_\mathrm{gas}$\\
				& (\MO\,yr$^{-1}$) & (10$^{-10}$ yr$^{-1}$) & (10$^{-10}$ yr$^{-1}$) & (10$^{-10}$ yr$^{-1}$) \\
		\hline
		NGC\,5394  & 4.9  & 8.0 & 10.2  & 10.0\\
		NGC\,5395  & 3.1  & 1.4  & 3.9 & 2.3\\
		NGC\,4567  & 0.67 & 4.3  & 5.2 & 4.5\\
		NGC\,4568  & 1.5  & 3.6 & 5.0  & 3.7\\
		UGC\,12914 & 1.6 & 1.0  & 1.6 & 1.0\\
		UGC\,12915 & 4.2 & 3.1  & 2.8 & 2.1\\
		NGC\,4038  & 8.4  & 14.8 & 13.5 & 9.7\\
		NGC\,4039  & 9.5  & 22.3 & 29.7 & 22.6\\
		\hline
	\end{tabular}}
	\label{derived_interact}
\end{table}

We derive two galaxy-scale SFEs, i.e., SFE$_\mathrm{mol}$ and SFE$_\mathrm{gas}$, defined as the following equations:
\begin{eqnarray}
	\label{deriveSFE}
	&& \mathrm{SFE}_\mathrm{mol} \ \mathrm{[yr^{-1}]} = \mathrm{SFR}/M_\mathrm{H_{2}},\\	
	\label{deriveSFEgas}
	&&\mathrm{SFE}_\mathrm{gas} \ \mathrm{[yr^{-1}]} = \mathrm{SFR}/(M_\mathrm{H_{2}}+M_\mathrm{H\emissiontype{I}}).
\end{eqnarray}
While all isolated galaxies have lower SFE$_\mathrm{gas}$ than 10$^{-9}$ yr$^{-1}$, two of eight interacting galaxies have SFE$_\mathrm{gas}$ higher than 10$^{-9}$ yr$^{-1}$.
However, the $p$-values of the Kolmogorov--Smirnov test for SFE$_\mathrm{mol}$ and SFE$_\mathrm{gas}$ are 0.42 and 0.52, respectively.
These results indicate that both SFE$_\mathrm{mol}$ and SFE$_\mathrm{gas}$ for interacting galaxies and isolated galaxies are obtained from the same parents, consistent with \citet{Casasola04}.
\citet{Young96} reported that global SFE$_\mathrm{gas}$ in strongly interacting galaxies is higher than those found in normal spiral galaxies.
The discrepancy might be caused by the increasing SFE as a function of the merger stage.

\begin{table}[tbp]
	\tbl{Galaxy-scale star formation properties of the isolated galaxies.}{%
	\begin{tabular}{ccccc}
		\hline
		Galaxy & SFR                     &    sSFR			 & SFE$_\mathrm{mol}$ & SFE$_\mathrm{gas}$\\
		       & (\MO\,yr$^{-1}$) & (10$^{-10}$ yr$^{-1}$) & (10$^{-10}$ yr$^{-1}$) & (10$^{-10}$ yr$^{-1}$) \\
		NGC\,253  & 2.5 & 6.3 & 15.6 & 9.6 \\
		NGC\,2903 & 1.0 & 3.4 & 6.3 & 2.6 \\
		NGC\,3184 & 0.6 & 5.6 & 5.5 & 2.0 \\
		NGC\,3351 & 0.9 & 2.6 & 8.2 & 3.9 \\
		NGC\,3521 & 1.1 & 2.4 & 3.8 & 1.7 \\
		NGC\,3627 & 3.1 & 2.9 & 3.5 & 3.1 \\
		NGC\,4736 & 0.4 & 1.1 & 10.0 & 5.7 \\
		NGC\,5055 & 1.1 & 2.1 & 3.8 & 1.1 \\
		NGC\,5236 & 3.3 & 6.4 & 16.5 & 8.9 \\
		NGC\,5457 & 2.7 & 6.9 & 8.4 & 1.6 \\
		NGC\,6946 & 2.9 & 5.8 & 7.3 & 3.8 \\
		\hline
	\end{tabular}}
	\label{derived_isolated}
\end{table}

All derived properties for the interacting galaxies and the isolated galaxies are summarised in tables \ref{derived_interact} and \ref{derived_isolated}.
Table \ref{Average} shows the mean galaxy-scale properties of interacting galaxies and the isolated galaxies.

\begin{table}[tbp]
	\tbl{Mean star formation properties of interacting galaxies and isolated galaxies.}{%
	\begin{tabular}{ccc}
		\hline
		Quantities\footnotemark[*] & Interacting Galaxies & Isolated Galaxies\\
		\hline
		Number	&	8	& 11	\\
		$M_\mathrm{H_{2}}$ [10$^{10}$ \MO]		& 8.0$\pm$7.0 & 3.4$\pm$2.9 \\
 		SFR [\MO\,yr$^{-1}$]			& 4.1$\pm$3.3 & 1.5$\pm$1.2	 \\
		sSFR [10$^{-10}$ yr$^{-1}$]		& 7.3$\pm$7.0 & 4.1$\pm$2.0  \\
		SFE$_\mathrm{mol}$	[10$^{-10}$ yr$^{-1}$]			& 9.0$\pm$8.7 & 8.1$\pm$4.3 \\		
		SFE$_\mathrm{gas}$	[10$^{-10}$ yr$^{-1}$]			& 7.0$\pm$6.7 & 4.0$\pm$2.8 \\
		\hline
	\end{tabular}}
	\label{Average}
	\begin{tabnote}
		\footnotemark[$*$]{The errors are the 1$\sigma$ uncertainties.}
	\end{tabnote}
\end{table}

\begin{figure}[tbp]
	\centering
	\includegraphics{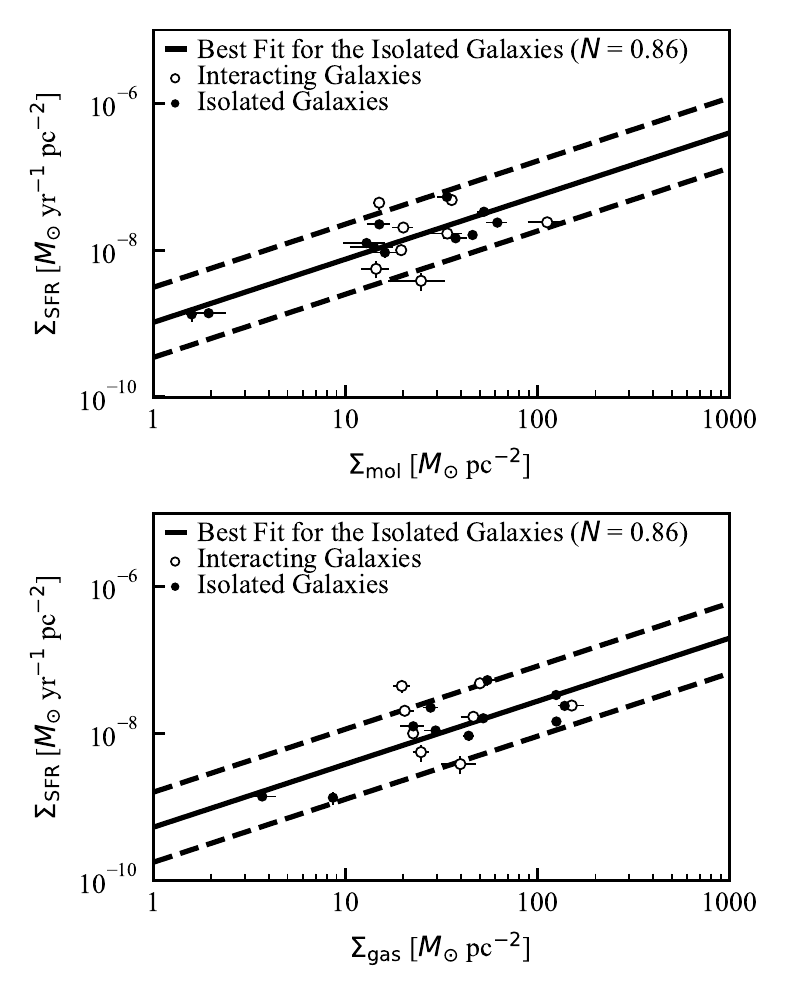}
	\caption{Global Kennicutt--Schmidt law. (Top) Relationship between the surface density of star formation rate and molecular hydrogen gas mass for the interacting galaxies and the isolated galaxies. (Bottom) Relationship between the surface density of star formation rate and total gas mass for the interacting galaxies and the isolated galaxies. The line is the best least-squares fit for all isolated galaxies, and the dashed lines represent SFR is enhanced and decreased by a factor of three from the best fit.}
	\label{globalSKlaw}
\end{figure}

A galaxy-scale relationship between the surface density of SFR and ISM contents, i.e., the Kennicutt--Schmidt law using $\Sigma_\mathrm{mol}$ is shown in figure \ref{globalSKlaw}(top).
The best-fitting power-law function for isolated galaxies is
\begin{equation}
	\mathrm{log} \ \Sigma_\mathrm{SFR} = -(8.98\pm0.17) + (0.86\pm0.13) \mathrm{log} \ \Sigma_\mathrm{mol},
\end{equation}
where $\Sigma_\mathrm{SFR}$ and $\Sigma_\mathrm{mol}$ are the surface density of SFR [\MO\,yr$^{-1}$ pc$^{-2}$] and H$_{2}$ mass [\MO\,pc$^{-2}$], respectively.
$\Sigma_\mathrm{SFR}$ and $\Sigma_\mathrm{mol}$ are derived assuming that all star-forming regions and molecular gas are within $R_\mathrm{K20}$.
Although the dynamic range is narrow and the number of galaxies is small, the best least-squares fit for the isolated galaxies is consistent with the results of a much larger sample from the COLD GASS survey, which shows the index of 0.86$\pm$0.21 \citep{Saintonge12}.
We found no clear difference between interacting galaxies and isolated galaxies in the Kennicutt--Schmidt law: seven out of eight interacting galaxies have not enhanced SFRs by more than a factor of 3 compared to the best-fit line for the isolated galaxies.

Figure \ref{globalSKlaw}(bottom) shows the Kennicutt--Schmidt law adopting $\Sigma_\mathrm{gas}$ instead of $\Sigma_\mathrm{mol}$.
The best least-squares fit for isolated galaxies is
\begin{equation}
	\mathrm{log} \ \Sigma_\mathrm{SFR}  = -(9.28\pm0.32) + (0.86\pm0.19) \mathrm{log} \ \Sigma_\mathrm{gas} ,
\end{equation}
where $\Sigma_\mathrm{gas}$ is the surface density of total gas mass.
The result is similar to the case of $\Sigma_\mathrm{SFR}-\Sigma_\mathrm{mol}$ relation: 
most interacting galaxies do not exceed the relation for isolated galaxies considering a scatter.
The exceptional galaxy, NGC\,4039, is the constituent galaxy of the Antennae Galaxies.
The active star formation in the Antennae Galaxies is consistent with the previous studies that some interacting galaxies in the mid stage already show violent star formation (e.g., \cite{Zhu03}).

In the galaxy-scale Kennicutt--Schmidt law for both $\Sigma_\mathrm{SFR}-\Sigma_\mathrm{mol}$ and $\Sigma_\mathrm{SFR}-\Sigma_\mathrm{gas}$, star formation activity of our sample interacting galaxies have almost as same as that in isolated galaxies.
Although several authors have reported a bi-modality in the Kennicutt--Schmidt law (sequences of starburst and discs) \citep{Daddi10, Genzel10, Violino18}, our sample interacting galaxies do not show the bi-modality.
The simulations performed by \citet{DiMatteo07} show that interacting galaxies are not in the sequence of starburst, while mergers have higher SFE than isolated galaxies.
Our result is in agreement with their simulations.

\section{Star formation rate and star formation efficiency on a kpc scale}
\label{LSF}
To understand how star formation is enhanced during a galaxy interaction, it is important to reveal when and where properties of star formation are changed.
As presented in section \ref{GSF}, our sample does not show enhanced star formation on a galaxy scale.
However, it does not mean there is no local enhancement of star formation activity in these galaxy pairs.
For this reason, we investigate kpc-scale star formation indicators (SFR, SFE$_\mathrm{mol}$, and SFE$_\mathrm{gas}$).

\subsection{Maps of star formation rate and star formation efficiency}
First, we calculate SFR using H$\alpha$ (or FUV for Arp\,84 and VV\,254) and MIPS 24 $\mu$m (or IRAC 8 $\mu$m for Arp\,84) data in a map base.
Since we will investigate not only SFR but also SFE$_\mathrm{mol}$ and SFE$_\mathrm{gas}$, these data are convolved into the same angular resolution as CO data (19$\farcs$3), which is the lowest among the data and re-gridded to 7$\farcs$5 per pixel.
The reason for adopting a smaller pixel size than the Nyquist sampling rate ($\sim$10$^{\prime\prime}$) is the limitation of the CO observations (the OTF method).
In the OTF method, the antenna is driven continuously in a region to be mapped, and the data are acquired in a short interval.
But antenna jitter and pointing accuracy do not make the data points align any regular grid.
Thus, the data should be re-gridded onto a regular (rectangular) grid adopting a gridding convolution function during data reduction.
This process makes the resultant beam size larger than the telescope beam (For the CO observations, whose rest frequency is 115.27 GHz, with the Nobeyama 45-m telescope, the telescope beam is \timeform{15''}) (see, figure 6 in \cite{Sawada08}).
In addition, the peak temperature of a point source becomes lower. 
For example, if one choose the pixel size of $\sim$0.8 $\times$ the original telescope beam, the data can be sampled with a Nyquist rate and the effective beam size becomes $\sim$1.6 larger than the original telescope beam.
However, such a large grid spacing leads to lower response to a point source (0.4 $\times$ the single-beam observation), which can not detect significant emission from some interacting galaxies.
On the other hand, the smaller pixel size leads to the effective integration time is smaller, and thus the noise level becomes larger with the square of the pixel size.
Considering these effects, we chose the pixel size of 7$\farcs$5, which is the Nyquist sampling rate of the original telescope beam, in this paper.
Note that this over-sampling leads to yield a better correlation in pixel-to-pixel comparison.
The effect of different sampling is discussed in Appendix.

On deriving SFR, we applied equations \ref{deriveSFR1} and \ref{deriveSFR2} on each pixel.
We note that maps of SFR of interacting galaxies are not corrected for the inclination angle since the map contains two galaxies with a different inclination and is hard to handle.
On the other hand, we correct the inclination angle effects when we plot the surface density of SFR in section \ref{resolvedkslaw}.

To investigate where active star formation occurs, we also derive SFE$_\mathrm{mol}$ and SFE$_\mathrm{gas}$ maps for interacting galaxies based on equations \ref{deriveSFE} and \ref{deriveSFEgas}.
Note that maps of SFE$_\mathrm{mol}$ and SFE$_\mathrm{gas}$ are not affected by the inclination because the inclination angle affects $\Sigma_\mathrm{SFR}$, $\Sigma_\mathrm{mol}$ and $\Sigma_\mathrm{gas}$ in the same manner.
If SFR and gas masses have uncertainties of 0.4 dex, as noted in section \ref{values}, the uncertainties of SFE$_\mathrm{mol}$ and SFE$_\mathrm{gas}$ are roughly 55 and 60\%, respectively.

Figures \ref{fig_SFR}, \ref{fig_SFEmol}, and \ref{fig_SFEgas} show maps of SFR, SFE$_\mathrm{mol}$, and SFE$_\mathrm{gas}$ with contours of CO integrated intensity for interacting galaxies.
Since the pointing accuracies of all observations are better than 5$^{\prime\prime}$ as described in section \ref{data}, the spatial variations of each map are reliable.
Therefore, we mask the values below 1 $\sigma$ noise error of the surface densities of SFR, molecular gas, and total gas in these maps.

\begin{figure*}
	\centering
	\includegraphics{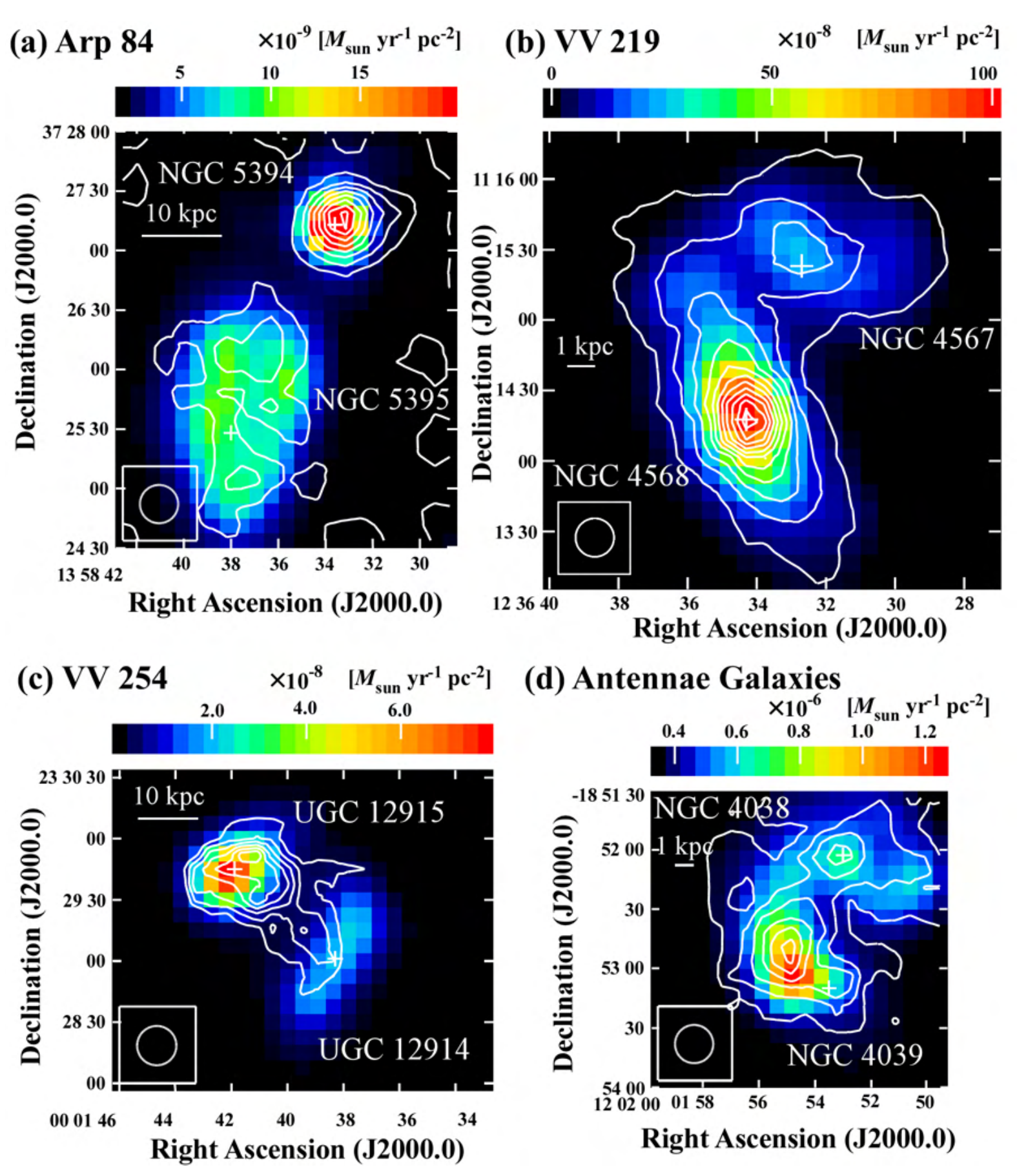}
	\caption{
		(a) A SFR image with CO integrated intensity contours for Arp\,84. The contours are 2.81 $\times$ 2, 3, 4, ... K km s$^{-1}$.
		(b) The same as (a) but for VV\,219. The contours are 7.20 $\times$ 1, 2, 3, ... K km s$^{-1}$.
		(c) The same as (a) but for VV\,254. The contours are 6.63 $\times$ 2, 3, 4, ... K km s$^{-1}$.
		(d) The same as (a) but for the Antennae Galaxies. The contours are 23.7 $\times$ 1, 2, 3, ... K km s$^{-1}$.
	}
	\label{fig_SFR}
\end{figure*}

\begin{figure*}
	\centering
	\includegraphics{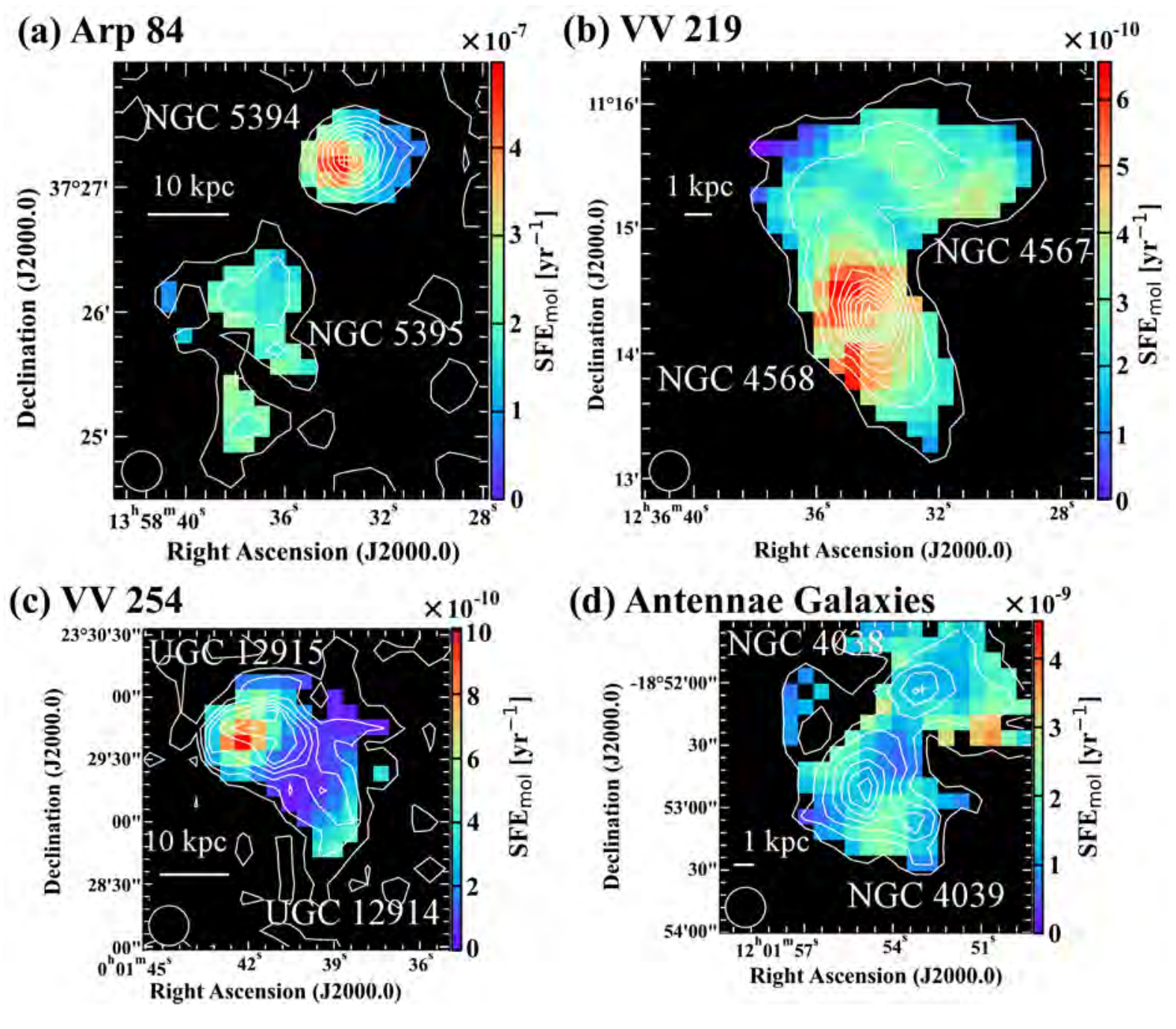}
	\caption{
		(a) The same as figure \ref{fig_SFR}(a) but a colour image is SFE$_\mathrm{mol}$.
		(b) The same as figure \ref{fig_SFR}(b) but a colour image is SFE$_\mathrm{mol}$.
		(c) The same as figure \ref{fig_SFR}(c) but a colour image is SFE$_\mathrm{mol}$.
		(d) The same as figure \ref{fig_SFR}(d) but a colour image is SFE$_\mathrm{mol}$.
	}
	\label{fig_SFEmol}
\end{figure*}
				
\begin{figure*}
	\centering
	\includegraphics{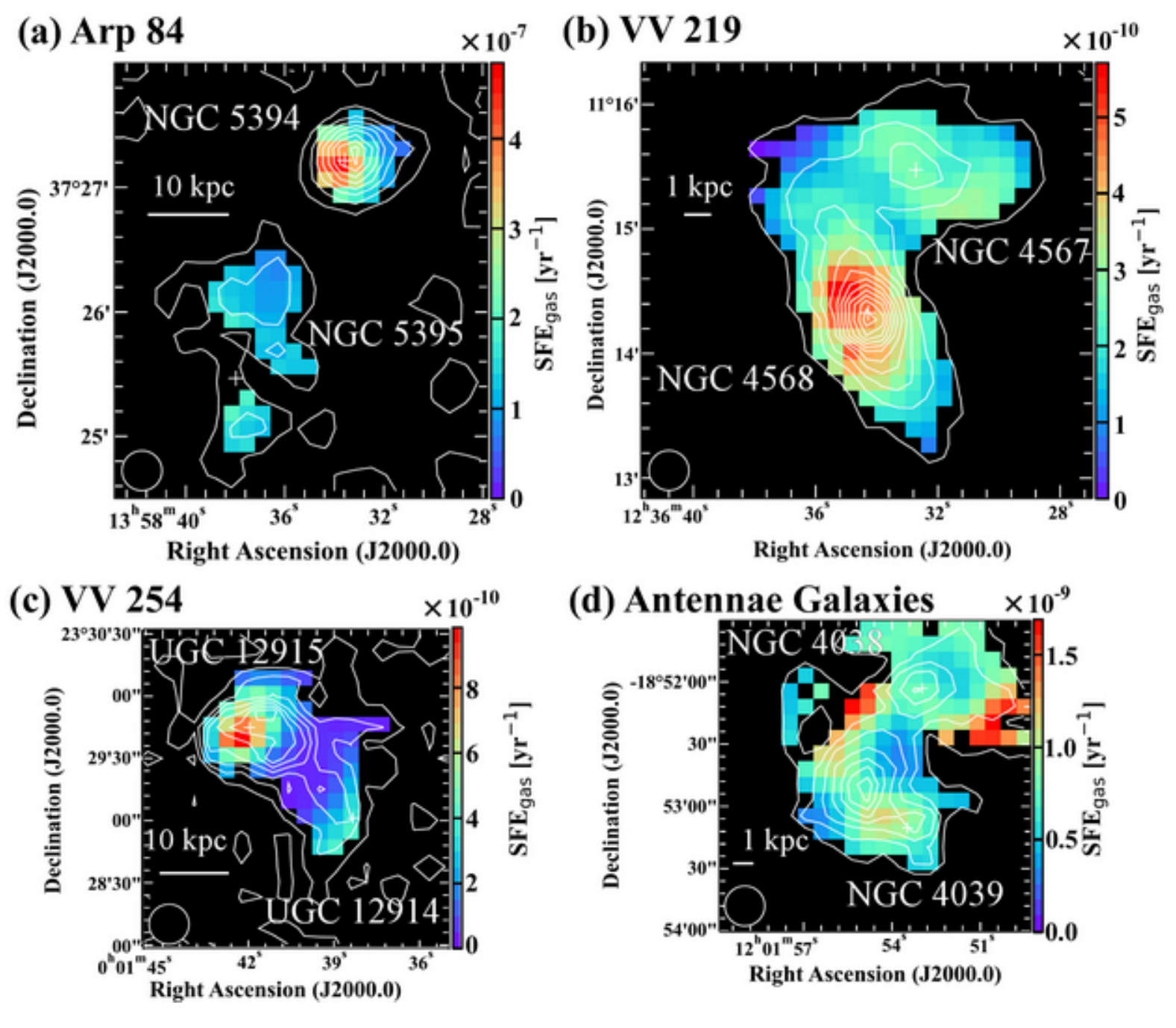}
	\caption{
		(a) The same as figure \ref{fig_SFR}(a) but a colour image is SFE$_\mathrm{gas}$.
		(b) The same as figure \ref{fig_SFR}(b) but a colour image is SFE$_\mathrm{gas}$.
		(c) The same as figure \ref{fig_SFR}(c) but a colour image is SFE$_\mathrm{gas}$.
		(d) The same as figure \ref{fig_SFR}(d) but a colour image is SFE$_\mathrm{gas}$.
	}
	\label{fig_SFEgas}
\end{figure*}

\subsection*{Arp\,84}
Figure \ref{fig_SFR}(a) shows that star formation activity in NGC\,5394 (top-right) is high compared with NGC\,5395 (bottom-left).
FUV and H$\alpha$ \citep{Roche15} show weak star-forming regions along the spiral arm and a circular-shaped active star-forming region in the galactic centre.
On the other hand, there is no star-forming region along the spiral arm in our SFR map.
This could be due to low angular resolution.
The peak of SFR, which is about 2.0 $\times$ 10$^{-8}$ \MO\,yr$^{-1}$ pc$^{-2}$, locates the centre of the galaxy, while the CO distribution is slightly lopsided toward the western side.
With higher resolution CO data  ($5\farcs5 \times 4\farcs3$), \citet{Iono05} also reported that the peak of CO is located westward from the centre of the galaxy.
This discrepancy in the distribution between CO and SFR is emphasized in SFE maps in figures \ref{fig_SFEmol}(a) and \ref{fig_SFEgas}.
The active star-forming region is slightly lopsided toward the nearer side to its companion, although the highest SFE is still at the centre of NGC\,5394.

NGC\,5395 shows the more complicated distribution of star-forming regions.
The distribution of SFR differs from that of molecular gas (CO) and old stars ($K_\mathrm{s}$) (see figure \ref{Ks}).
A peak of SFR locates near the centre of NGC\,5395 but \timeform{10''} (2.5 kpc) to the north-east.
Star-forming regions mainly trace the distribution of the old stars.
The peak of SFR, which is found at the end of the tidal tail in CO, does not coincide with the peaks of CO.

NGC\,5395 has two local peaks in both SFE$_\mathrm{mol}$ and SFE$_\mathrm{gas}$, which correspond to the peaks of SFR described above.
The local peak of SFE$_\mathrm{mol}$ and SFE$_\mathrm{gas}$ at the tidal tail of NGC\,5395 has comparable values to those at the centre of NGC\,5395.
The edge of the tidal tail, where CO gas is most abundant in NGC\,5395, shows the lowest SFE$_\mathrm{mol}$ and SFE$_\mathrm{gas}$ in NGC\,5395, even considering the uncertainty.
This fact suggests active star formation with high SFEs cannot be caused only by the accumulation of molecular gas due to interaction.

\subsection*{VV\,219}
As illustrated in figure \ref{fig_SFR}(b), the most intense star-forming activity in NGC\,4568 (bottom-left) is at its centre, and SFR gradually decreases as the radius increases.
These features are also seen in isolated spiral galaxies.
The active star formation of NGC\,4567 (top-right) takes place along its spiral arms.
The overlap region shows slightly enhanced star formation, and SFR is comparable to that in the disc of NGC\,4567.

Figures \ref{fig_SFEmol}(b) and \ref{fig_SFEgas}(b) show the distributions of SFE$_\mathrm{mol}$ and SFE$_\mathrm{gas}$ in VV\,219.
The spiral arms in NGC\,4567 show higher SFEs.
In particular, the southwestern arm has the highest SFEs.
Another region with high SFEs is seen near the centre of NGC\,4568, as often seen in spiral galaxies. 
The overlap region also shows a slight enhancement of SFEs, although it is insignificant when considering the uncertainty.
However, high angular resolution CO image (\timeform{2''}$\times$\timeform{2''}) shows H$\alpha$ blobs are found where CO emission is weak (figure 3 in \cite{Kaneko18}).
Therefore, SFEs are actually enhanced in the overlap region.

\subsection*{VV\,254}
Figures \ref{fig_SFR}(c), \ref{fig_SFEmol}(c), and \ref{fig_SFEgas}(c) are the maps of SFR, SFE$_\mathrm{mol}$ and SFE$_\mathrm{gas}$ of VV\,254, respectively.
The VV\,254 system also shows relatively simple distributions of SFR and SFEs like the VV\,219 system.
High star formation activities are seen in the centre of UGC\,12914 (bottom-right) and UGC\,12915 (top-left),
although UGC\,12915 is a more active star-forming galaxy.
Additionally, a region showing high SFR in UGC\,12915 is slightly extended toward the overlap region, corresponding to the star-forming region previously reported by \citet{Komugi12}.
Star formation in UGC\,12914 is associated with the warped stellar disc.
Star formation is not active in the overlap region despite a large amount of molecular gas and atomic gas.

Unlike SFR, the highest SFE$_\mathrm{mol}$ and SFE$_\mathrm{gas}$ locate at $<$ \timeform{10''} to the south from the centre of UGC\,12915.
This region also does not correspond to the peak of CO.
Higher angular-resolution CO data obtained by \citet{Iono05} shows a giant molecular cloud apart from the disc of UGC\,12915. 
The position of this cloud matches the location of the highest SFE$_\mathrm{mol}$ and SFE$_\mathrm{gas}$ of VV\,254.
Therefore, the high SFE$_\mathrm{mol}$ and SFE$_\mathrm{gas}$ may be due to the \label{key}active star formation of this cloud.
In UGC\,12914, SFE$_\mathrm{mol}$ and SFE$_\mathrm{gas}$ in the warped part of the disc (the northern disc) are higher than in a non-warped part.
However, due to the low CO sensitivity, we cannot see the distributions of SFEs in the outer disc of UGC\,12914 in this study.

\subsection*{The Antennae Galaxies}
Figures \ref{fig_SFR}(d), \ref{fig_SFEmol}(d), and \ref{fig_SFEgas}(d) reveal that star-forming regions are embedded along the stellar tidal arms of the Antennae Galaxies.
A peak of SFR in NGC\,4038 (top) is at its centre, while a peak of NGC\,4039 (bottom) is off the centre.
Besides, most intense star formation occurs at the southern overlap region reaching as high as $\sim$10$^{-6}$ \MO\,yr$^{-1}$ pc$^{-2}$
where supergiant molecular complexes ($5\times10^{5}-9\times10^{8} \MO$) \citep{Wilson00} and superstar clusters ($>10^{5} \MO$) \citep{Whitmore99, ZF99} have been found.
Although the overlap region shows higher star formation activity compared to the other regions with rich CO, 
the peak of SFR does not coincide with the peak of CO as other interacting galaxies.

The maps of SFEs (figure \ref{fig_SFEmol}(d)) show that the most efficient star formation is seen at the southern overlap region, which corresponds to the peak of SFR and the SGMC\,4-5 in \citet{Wilson00},
 while SFE in the peak of CO in the overlap region, which is the SGMC\,1, is not so high.
The second most efficient star formation is seen at a branching point of the southern tidal arm of NGC\,4038.
The central regions of both galaxies undergo relatively calm star formation.
Even in the calm star-forming region, SFE is one order higher than other interacting galaxies.
We note that figure \ref{fig_SFEgas}(d) shows the highest SFE$_\mathrm{mol}$ at the branching point of the tidal arm of NGC\,4038 and the overlap region.
In this region, a strong [N\emissiontype{II}] line, suggesting massive star formation, is detected.
On the other hand, CO is just above 1 $\sigma$ detection level in our single-dish data, and the interferometric CO image shows no CO emission \citep{Wilson00}.
These facts imply that molecular gas is few and diffuse in this region.

\subsection{Causes of an enhancement of star formation efficiency}
In isolated spiral galaxies, SFE is almost constant where H$_{2}$ is dominant in the ISM (the inner part of galaxies) and decreases with increasing galactic radius \citep{RY99,Leroy08}.
Barred galaxies tend to show local SFE peaks at their bar-end \citep{Muraoka09,Maeda20}.
SFE of spiral galaxies has typically symmetric distribution.

On the other hand, we found local (asymmetric and off-centre) enhancement of SFE in interacting galaxies.
Although the number of targets is too small to show a unified enhancement mechanism, it would be meaningful to investigate reasons for local enhancement.
Here we discuss the cause of the enhancement of SFE of our targets.
The main topics are: (1) a possibility of environmental effects by a galaxy cluster rather than galaxy interaction, (2) a cause of high SFE in the secondary galaxy of a minor merger, and (3) a trigger of higher SFE at the collision front.

\subsubsection{Ram pressure compression in a galaxy cluster}
First, we examine the possibility of high SFE in NGC\,4567, which constructs the VV\,219 pair, due to the environmental effect from a cluster.
VV\,219 is located at a projected distance of $\sim$1 Mpc from the centre of Virgo Cluster.
NGC\,4567 is known as an H\emissiontype{I}-deficient galaxy \citep{Gavazzi05}.
These facts imply that NGC\,4567 may be suffered from ram pressure from the cluster.
Ram pressure can strip H\emissiontype{I} gas in the outer disc and compress the remaining H\emissiontype{I} gas in H\emissiontype{I}-deficient galaxies.
\citet{KY89} compared thermal, magnetic, turbulent, and gravitational energy density with ram pressure for molecular clouds and showed that
ram pressure might be one order of magnitude less than gravitational and turbulent energy densities.
Since turbulent and magnetic forces balance gravity in molecular clouds in their estimate,
ram pressure is not important to compress molecular clouds.
Thus, we expect that enhancement on star formation activity is not caused by ram pressure based on their scenario.

On the other hand, \citet{BC03} demonstrated using hydrodynamical simulations that ram pressure compresses a self-gravitating cloud and triggers star formation in a cloud.
Some galaxies show star formation triggered by ram pressure in Virgo Cluster \citep{KK04,Cramer20}. 
However, the fact that star-forming regions triggered by ram pressure are concentrated on the edge of the galactic disc in these galaxies implies that an influence of ram pressure is local.
If the ram pressure is perpendicular to the disc of NGC\,4567, gas compression should happen on the whole disc.
In that case, SFE may get equally higher in the disc by ram pressure.
However, it is difficult to explain high SFE in the disc of NGC\,4567 only with the ram pressure since spiral arms of NGC\,4567 show higher SFE.
We conclude that ram pressure is not the main trigger of high SFE in the disc of NGC\,4567.

\subsubsection{An effect of galaxy mass ratio}
Next, we investigate a reason for high SFE in the secondary galaxy.
Arp\,84 and VV\,219 show higher SFEs are found in the secondary galaxy (i.e., NGC\,5394 and NGC\,4567) of the pair. 
Several studies have pointed out that the influence of the mass ratio of galaxies is important for star formation history during the interaction.
More active star formation is reported in a secondary galaxy of minor interacting pairs \citep{Li08, WG07}.
Simulations have shown that galaxy interactions transfer angular momentum from gas to stars, which moves gas to the centre of progenitors \citep{BH96}.
Therefore, these cases can be interpreted that the primary galaxy is not tidally disturbed enough to induce radial gas infall that enhances star formation activity.
In contrast, the secondary galaxy experiences a significant tidal force.

To determine if this effect predominates in the early stage, we calculate the dynamical mass ratio of the target interacting galaxies.
Although the dynamical mass ratios of NGC\,5395 and NGC\,5394, and NGC\,4568 and NGC\,4567 are estimated as $\sim$100 and $\sim$10, respectively, according to \citet{Iono05},
they did not correct the inclination in deriving the dynamical mass.
We re-calculate the dynamical mass considering the inclination of galaxies.
The dynamical mass can be derived through the equation below:
\begin{equation}
	M_\mathrm{dyn}\ [\MO] = 2.32 \times 10^{5} \left( \frac{R_\mathrm{edge}}{\mathrm{kpc}}\right) \left(\frac{V_\mathrm{corr}}{\mathrm{km \ s^{-1}}} \right)^{2} ,
\end{equation}
where $R_\mathrm{edge}$ is the maximum radius where emission is detected above 3 $\sigma$ significance level, and $V_\mathrm{corr}$ (= $V_\mathrm{rot}$/sin $i$) is inclination corrected radial velocity.
$V_\mathrm{rot}$ is derived by $|V_\mathrm{obs}(R_\mathrm{edge})-V_\mathrm{sys}|$, where $V_\mathrm{obs}(R_\mathrm{edge})$ is the observed velocity at $R_\mathrm{edge}$, and $V_\mathrm{sys}$ is the systemic velocity.
We calculate the dynamical mass from both CO and H\emissiontype{I} data and summarise it in table \ref{dynamical_mass}.
\begin{table*}[tbp]
	\tbl{The dynamical mass of targets.}{%
	\begin{tabular}{ccccc}
		\hline
		       & \multicolumn{2}{c}{Dynamical Mass [\MO]} & \multicolumn{2}{c}{Mass Ratio} \\
		Galaxy & H\emissiontype{I} & CO & H\emissiontype{I} & CO  \\
		\hline
		NGC\,5394 & $>$2.4$\times$10$^{9}$ & $>$9.4$\times$10$^{9}$  & & \\
		NGC\,5395 & 4.5$\times$10$^{11}$ & 2.8$\times$10$^{11}$ & $<$188 & $<$30 \\
		\hline
		NGC\,4567 & 2.1$\times$10$^{10}$ & 2.2$\times$10$^{10}$ &  & \\
		NGC\,4568 & 9.4$\times$10$^{10}$ & 8.1$\times$10$^{10}$ & 4.5 & 3.7 \\
		\hline
		UGC\,12914 & 6.3$\times$10$^{11}$ & 2.7$\times$10$^{10}$ & & \\
		UGC\,12915 & 2.7$\times$10$^{11}$ & 2.7$\times$10$^{10}$ & 2.3 & 1.0 \\
		\hline
		The Antennae Galaxies & --- & --- & --- & ---\\
		\hline
	\end{tabular}}
	\label{dynamical_mass}
\end{table*}
Because of its nearly face-on disc of NGC\,5394, a mass ratio for Arp\,84 has a large uncertainty.
Therefore, we adopt the mass ratio of Arp\,84 from the numerical simulation by \citet{Kaufman99}.
They assumed that NGC\,5395 is 4 times heavier than NGC\,5394 (i.e., the mass ratio of 4) and successfully reproduced the morphology of Arp\,84.
If we use this mass ratio of 4, Arp\,84 is a minor merger (a mass ratio $>$ 3).
Since VV\,219 has a mass ratio larger than 3 estimated from both CO and H\emissiontype{I} data, VV\,219 is also a minor merger. 
We conclude that Arp\,84 and VV\,219 can be regarded as an early phase of a minor merger.
Based on the unequal-mass galaxy merger simulations (mass ratio of from 2.3 to $>$ 20) \citep{Cox08}, star formation activity in the secondary galaxy is significantly affected by gravitational perturbation.
Although it makes a small contribution to the overall SFR of the pair, the enhancement of SFR in the secondary galaxy could be larger than that in the primary galaxy.
This suggests that SFE in the secondary galaxy is higher than that in the primary galaxy.

However, the distributions of star-forming regions may be different from simulations.
In the simulations, star formation occurs in the gas concentrated into the central region by an interaction (e.g., \cite{BH96}; \cite{TCB10}).
On the other hand, star-forming regions of NGC\,5394 and NGC\,4567 are distributed over the whole disc (H$\alpha$ emission is seen not only in the central region of NGC\,5394 but also in the edge of the northern tidal arm, which is slightly outside the field of view of CO \citep{Roche15}).
The central concentration of molecular gas is still low compared with that of stars in these galaxies (see Paper I).

We have suggested that largely prevailed shocks throughout the disc of galaxies in Paper II.
\citet{Roche15} indicated that high [N\,\emissiontype{II}]/H$\alpha$ and other line ratios of the outer region of NGC\,5394 are explained as a composite of shock and H\,\emissiontype{II} region.
Similarly, other interacting galaxies also show the off-centre and widely-spread shock (e.g., \cite{RKD11}; \cite{Wild14}; \cite{Saito15}).
If a shock occurs during the interaction, the shock in secondary galaxies should be stronger than that in primary galaxies due to the relative strength of a gravitational disturbance.
Gas compressed by the shock leads to form stars efficiently.
Hence, our finding that secondary galaxies show higher SFE spreading over the whole discs is consistent with the existence of such shocks.
Other pairs in an early phase of minor tidal interaction also show active star formation in the disc of secondary galaxies, as shown in \citet{KK04}.
This fact implies that an enhancement of star formation activity in the galactic disc of a secondary galaxy seems to be a rather general phenomenon in the early phase of the minor merger. 
Although the disturbance of the velocity field, which indicates shocks, was not found with our resolution in the discs of NGC\,4567,
a shock tracer observation with higher angular resolution may reveal direct evidence of the shocks.

\subsubsection{High SFE at a collision front}
\label{mechanism}
We consider the fact that a slight enhancement of SFEs is seen at the collision interface of Arp\,84 and VV\,219.
This can be naturally explained by star formation triggered by the compression of the gas at the interface.
High-resolution simulations by \citet{Saitoh09} reproduced the shock-induced starburst at the interface.
They showed that a giant filament of ISM is formed at the interface, and then starburst is induced there.
\citet{Kaneko18} clearly showed the large filamentary molecular collision front in the overlap region of VV\,219.
Although the star formation at the collision interface of Arp\,84 and VV\,219 is not as strong as demonstrated by \citet{Saitoh09},
there are a few large star-forming regions in the interface \citep{Koopmann01,Kaufman99}.
These star-forming regions have higher SFEs, as shown in figures \ref{fig_SFEmol}(b) and \ref{fig_SFEgas}(b).
Therefore, such star-forming regions at the collision front may be at the onset phase of starburst.
Supergiant molecular complexes and superstar clusters have been found in the Antennae Galaxies \citep{Whitmore99,Wilson03},
which are in the more advanced stage of the interaction and show higher star-forming activity than other pairs.
The shock-induced star formation is the cause of the enhancement of SFE at the colliding regions. 

What makes a difference in star formation activity at the overlap region among Arp\,84, VV\,219, and VV\,254?
Although these pairs possess plenty of molecular gas at their colliding region, only VV\,254 does not show enhanced star formation activity there.
We focus on a giant molecular cloud (GMC) collision during an interaction process.
When GMC-GMC collisions happen in the overlap region, GMCs are expected to be ionised.
It takes a long time, about 10$^{7}$ years, to reproduce of GMCs according to the scenario proposed by \citet{Braine04}.
After the GMC reproduction, dense gas would be formed in the GMCs.
Once dense gas is formed in re-formed GMCs, star formation would occur co-instantaneously.
Off-course, The dense gas formation time scale after ionisation of GMCs requires longer than the GMC reproduction time scale of 10$^{7}$ years.
However, the collision age of VV\,254 (2 $\times$ 10$^{7}$ years) is comparable to the GMC reproduction time scale.
In case that a face-on collision of VV\,254 ionises most GMCs in discs of progenitors, the star formation activity becomes depressed until dense molecular gas is formed in GMCs.
From figures \ref{fig_SFEmol}(c) and \ref{fig_SFEgas}(c), we can see that star formation activity is low even existing rich molecular gas with high molecular gas fraction in the overlap region of VV\,254.
Although constituent galaxies of Arp\,84 and VV\,219 are still close, the time after the collision of these pairs might be longer than the dense gas formation time scale.
If that is the case, the difference of star formation activity at the overlap region among Arp\,84, VV\,219, and VV\,254 is attributed to the time after the collision.

Another way to explain the difference of SFE at the collision interface could be the difference in the relative velocity of two galaxies.
According to the discussion about H\emissiontype{I} gas by \citet{Condon93},
the relative velocity of UGC\,12914 and UGC\,12915 is about 600 km s$^{-1}$.
The relative velocities of progenitors for Arp\,84 and VV\,219 are only 100 km s$^{-1}$ and 50 km s$^{-1}$, respectively.
When the relative velocity is large, the GMCs are ionised due to the collision.
Once the gas is ionised, it takes a long time to reproduce the GMC again.
This effect decreases the SFE just after the collision.
Thus, the collision front of interacting galaxies with high relative velocity like a prograde-prograde edge-on collision (e.g., VV\,254 pair) 
may show active star formation with low SFE.

\section{Resolved Kennicutt--Schmidt law}
\label{resolvedkslaw}
We examine the spatially resolved Kennicutt--Schmidt law in this section.
As previously stated, we find complex distributions in SFR and SFEs, and the offset between peaks of CO and SFR.
Thus, it is expected that the spatially resolved Kennicutt--Schmidt law for interacting galaxies might differ from that for isolated galaxies.

In this investigation, we consider the spatial resolution of the data.
\citet{Onodera10} showed this relationship would break with higher spatial resolution than $\sim$100 pc.
They attribute the breakdown of the law to the difference in the evolutionary stages of individual GMCs.
\citet{Bigiel08} also examined the SFR-gas relationship with different spatial resolutions.
They found that the index $N$, the coefficient $A$, and the scatter are weakly correlated with a spatial resolution from 200 pc to 1 kpc, which could be attributed to stellar feedback and cloud formation on the scale $\lesssim$ 300 pc.
If the stellar feedback and cloud formation are the main reason for changing the shape of the Kennicutt--Schmidt law, it is considered that index, the coefficient and the scatter are not affected on scales larger than 1 kpc.
Therefore, we convolve the data to the spatial resolution of 1 kpc for the isolated spiral galaxies since the original spatial resolutions of these data for the isolated spiral galaxies are 400--700 pc.
Since the spatial resolution of interacting galaxies is larger than 1 kpc, we do not convolve the resolution of these data.
For this operation, the index and the coefficient can be compared between the interacting galaxies and the isolated galaxies.
Note that the scatter of the interacting galaxies may be smaller than that in the isolated galaxies due to the larger spatial resolution.

For the analysis, we only use the data where CO emissions are detected higher than 3 $\sigma$ noise level after convolving the data for the investigation.

\subsection{Tracers of the surface density of ISM}
\begin{figure}[tbp]
	\centering
	\includegraphics{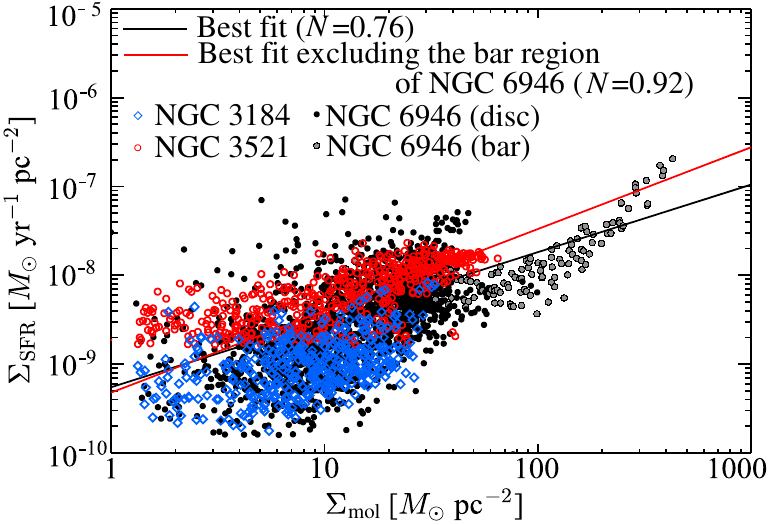}
	\caption{Star formation rate per unit area ($\Sigma_\mathrm{SFR}$) versus surface density of molecular gas ($\Sigma_\mathrm{mol}$) for isolated galaxies. 
		The black line indicates the best least-squares fit to whole data.
		The red line is the same as the black line but removing the data points of the bar region in NGC\,6946.}
	\label{SchmidtCOatlasMol}
\end{figure}

It has been discussed which value should be used as $\Sigma_\mathrm{ISM}$ in equation \ref{SKlaw}, namely, the surface density of total gas or molecular gas \citep{Bigiel08,WB02}.
This problem concerns an important issue what governs the physical process of conversion from interstellar gas to stars.
For this reason, in the beginning, we make two plots using isolated galaxy samples to check which relationship is a principle. 
One is the relationship between $\Sigma_\mathrm{SFR}$ and the surface density of molecular gas $\Sigma_\mathrm{mol}$, and another is between $\Sigma_\mathrm{SFR}$ and the surface density of total gas $\Sigma_\mathrm{gas}$.
In this investigation, NGC\,3184, NGC\,3521 and NGC\,6946 are used as isolated galaxy samples.
All fitting results are summarised in tables \ref{index_coefficient_mol} and \ref{index_coefficient_gas}.

\begin{table*}[tbp]
	\tbl{The fitting results for resolved $\Sigma_\mathrm{SFR}-\Sigma_\mathrm{mol}$ relation.}{%
		\begin{tabular}{cccc}
			\hline
			Sample & Index \itshape{N} & Coefficient \itshape{A}  & Coefficient of determination $R^{2}$ \\
			\hline
			Isolated Galaxies & 0.76$\pm$0.02 & -9.26$\pm$0.02 & 0.24 \\
			Isolated Galaxies (masked bar regions) & 0.92$\pm$0.02 & -9.24$\pm$0.02 & 0.27 \\
			Interacting Galaxies\footnotemark[$*$] &  1.66$\pm$0.03 & -10.5$\pm$0.05 & 0.86\\
			Interacting Galaxies (removed Antennae Galaxies) & 1.09$\pm$0.05 & -9.75$\pm$0.07 & 0.51 \\
			Interacting Galaxies (removed Antennae Galaxies \& VV\,254) & 1.16$\pm$0.03 & -9.72$\pm$0.04 & 0.85 \\
			\hline
	\end{tabular}}
	\label{index_coefficient_mol}
	\begin{tabnote}
		\footnotemark[$*$]The inclination correction for data points of Anttennae Galaxies is not performed. 
	\end{tabnote}
\end{table*}

\begin{table*}[tbp]
	\tbl{The fitting results for resolved $\Sigma_\mathrm{SFR}-\Sigma_\mathrm{gas}$ relation.}{%
		\begin{tabular}{cccc}
			\hline
			Sample & Index \itshape{N} & Coefficient \itshape{A}  & Coefficient of determination $R^{2}$ \\
			\hline
			Isolated Galaxies & 1.23$\pm$0.02  & -10.1$\pm$0.02  & 0.39 \\
			Isolated Galaxies (masked bar regions) & 1.33$\pm$0.03 & -10.2$\pm$0.03 &  0.46 \\
			Interacting Galaxies\footnotemark[$*$] & 1.84$\pm$0.03 & -10.9$\pm$0.05 & 0.88\\
			Interacting Galaxies (removed Antennae Galaxies) & 1.28$\pm$0.06 & -10.2$\pm$0.08 &  0.57 \\
			Interacting Galaxies (removed Antennae Galaxies \& VV\,254) & 1.29$\pm$0.03 & -10.1$\pm$0.05 & 0.85 \\
			\hline
	\end{tabular}}
	\label{index_coefficient_gas}
	\begin{tabnote}
		\footnotemark[$*$]The inclination correction for data points of Antennae Galaxies is not performed. 
	\end{tabnote}
\end{table*}

Figure \ref{SchmidtCOatlasMol} represents SFR per unit area ($\Sigma_\mathrm{SFR}$) versus surface density of molecular gas ($\Sigma_\mathrm{mol}$) for isolated galaxies.
The black line indicates the best least-squares fit to whole data with an index of 0.76$\pm$0.02.
An odd branch appears around $\Sigma_\mathrm{mol} > $ 40 \MO\,pc$^{-2}$ of NGC\,6946.
NGC\,6946 is a barred galaxy, and the bar region tends to have low SFE (e.g., \cite{Momose10,Hirota14,Maeda20}).
We performed an elliptical fit on the bar region of NGC\,6946 using an optical image.
We found that most points in this branch correspond to the bar region of NGC\,6946.
Therefore, we fitted the data again after masking the bar region of NGC\,6496.
The resultant red line is an index of 0.92$\pm$0.02 with $R^{2}$ of 0.27.

\begin{figure}[tbp]
	\centering
	\includegraphics{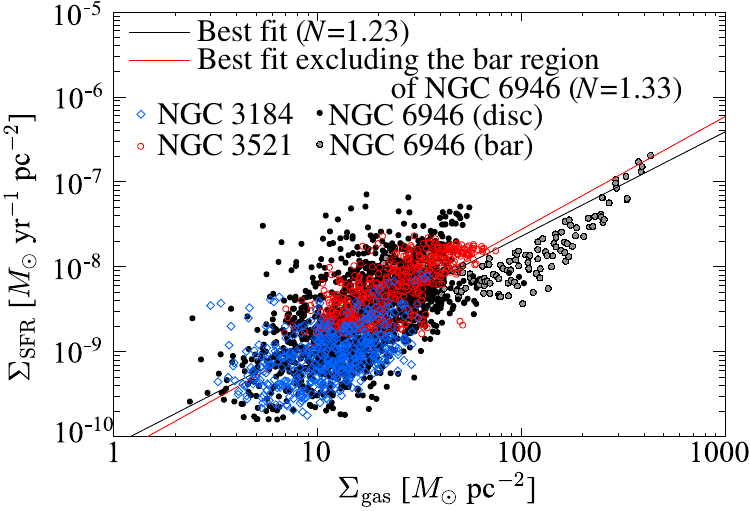}
	\caption{The same as figure \ref{SchmidtCOatlasMol} but for star formation rate per unit area ($\Sigma_\mathrm{SFR}$) versus surface density of total gas ($\Sigma_\mathrm{gas}$) for isolated galaxies.}
	\label{SchmidtCOatlas}
\end{figure}

Then, we made the same plot for the surface density of total gas ($\Sigma_\mathrm{gas}$) instead of molecular gas.
The result is shown in figure \ref{SchmidtCOatlas}.
The best least-squares fit to whole data has an index of 1.23$\pm$0.02.
Since the same branch due to the bar region of NGC\,6946 is found, we mask the bar region of NGC\,6496 again and find the index becomes 1.33$\pm$0.03.
This index is higher than those derived using $\Sigma_\mathrm{mol}$ as previously reported by \citet{Bigiel08} but is consistent with the value of the previous studies \citep{Kennicutt07}.
It may be due to the difference of a tracer of molecular gas.
\citet{Bigiel08} used the CO($J$ = 2--1) line for the tracer of molecular gas, assuming a constant CO($J$ = 2--1)/CO($J$ = 1--0) line ratio.
However, a recent study by \citet{Yajima21} illustrates that the CO($J$ = 2--1)/CO($J$ = 1--0) line ratio differs both galaxy-to-galaxy and within a galaxy.
They also point out that the index $N$ of the Kennicutt--Schmidt law gets smaller when using a constant CO($J$ = 2--1)/CO($J$ = 1--0) line ratio.

We find that the relation between $\Sigma_\mathrm{SFR}$ and $\Sigma_\mathrm{gas}$ gives a slightly better correlation ($R^{2}$ = 0.46) than that between $\Sigma_\mathrm{SFR}$ and $\Sigma_\mathrm{mol}$.
This result is inconsistent with \citet{Bigiel08}.
They attribute it to that star formation is more directly connected with molecular gas, at least on a kpc scale, since H\emissiontype{I} gas is saturated around $\Sigma_\mathrm{ISM} \simeq$ 10 \MO\,pc$^{-2}$.
Our result implies that H\emissiontype{I} gas also contributes to star formation, as discussed by \citet{Fukui19}.

\subsection{Resolved Kennicutt--Schmidt law for interacting galaxies}
\begin{figure}[tbp]
	\centering
	\includegraphics{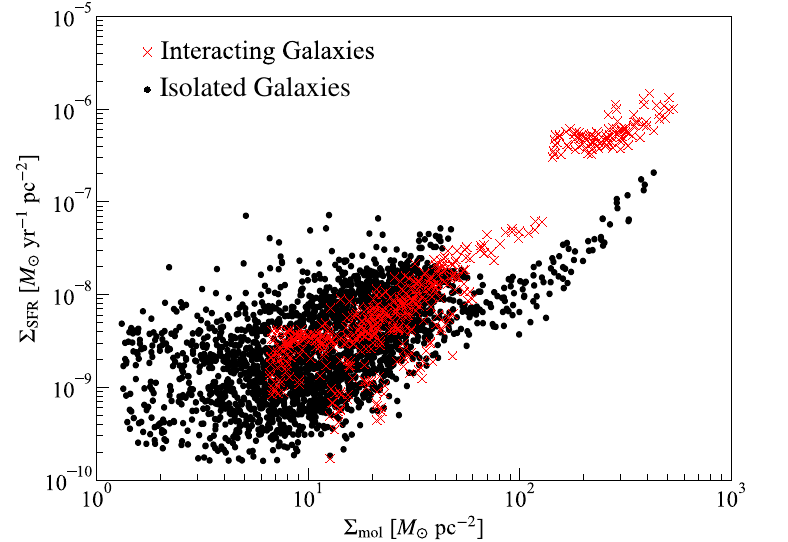}
	\caption{A relationship of $\Sigma_\mathrm{SFR}$-$\Sigma_\mathrm{mol}$ for all interacting galaxies (red crosses) and isolated galaxies (black dots).}
	\label{SchmidtMolAll}
\end{figure}

Figure \ref{SchmidtMolAll} shows a relationship between $\Sigma_\mathrm{SFR}$ and $\Sigma_\mathrm{mol}$ for interacting galaxies (red crosses) and isolated galaxies (black filled circles).
In a pixel-to-pixel comparison, interacting galaxies have a similar $\Sigma_\mathrm{SFR}$-$\Sigma_\mathrm{mol}$ relation to the isolated galaxies.

\begin{figure}[tbp]
	\centering
	\includegraphics{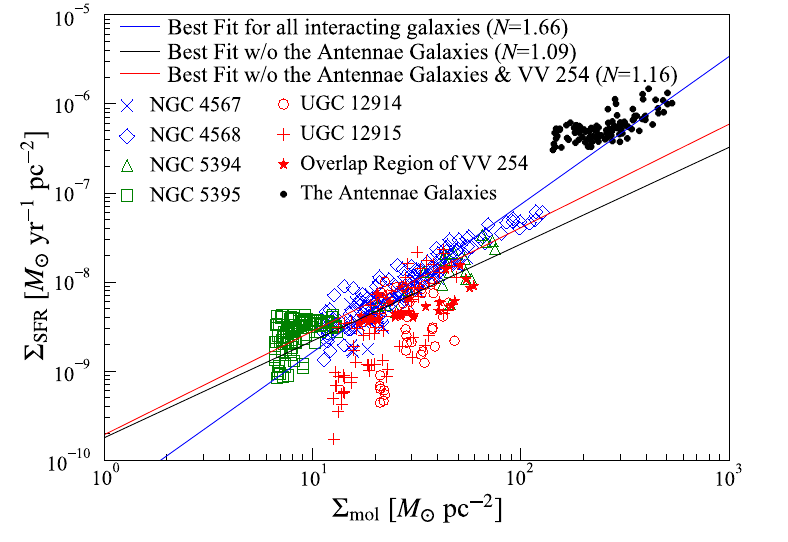}
	\caption{A relationship of $\Sigma_\mathrm{SFR}$-$\Sigma_\mathrm{mol}$ compiling all interacting galaxies. 
		The blue line represents the best least-squares fit to whole interacting galaxies.
		The black line indicates the best least-squares fit to whole interacting galaxies except for the Antennae Galaxies.
		The red line is the same as the black line but out of the Antennae galaxies and VV\,254.}
	\label{SchmidtMol}
\end{figure}

We also plot each progenitor of our interacting galaxies with different symbols to see the contribution to the relation (figure \ref{SchmidtMol}).
Even for interacting galaxies, there still exists a power-law relation with an index of 1.66$\pm$0.03.

We can see all points belonging to the Antennae Galaxies lie in the top-right region in the $\Sigma_\mathrm{SFR}$-$\Sigma_\mathrm{mol}$ plane.
However, due to their severely disturbed morphology,  we do not correct the inclination effect on $\Sigma_\mathrm{SFR}$ and $\Sigma_\mathrm{mol}$ for the Antennae Galaxies.
The inclination correction is important about the Kennicutt--Schmidt law since the inclination correction reduces both $\Sigma_\mathrm{SFR}$ and $\Sigma_\mathrm{mol}$.
Therefore, we remove the data of the Antennae Galaxies.
The index of the relation becomes 1.09$\pm$0.05.
This investigation corresponds to the Kennicutt-Schmidt law for only interacting galaxies in the early stage.

Suppose the inclination correction can be done on the Antennae Galaxies.
In that case, the index must become larger because the inclination-corrected data from the Antennae Galaxies move to the bottom left along with the index of 1, which is higher than the result of the original fit ($N$ = 1.09$\pm$0.05).

In figure \ref{SchmidtMol}, VV\,254 (red points) show slightly different properties from other galaxies.
Most data points of VV\,254 exist below the fitted line of isolated spiral galaxies (Some regions have one order of magnitude lower SFR than the best-fitted line for the isolated galaxies).
If we remove the data of the Antennae Galaxies and VV\,254 from the power-law fitting, we find the tight relation ($R^{2}$ = 0.85):
\begin{equation}
	{\rm log \ \Sigma_{SFR} = -(9.72\pm0.04) + (1.16\pm0.03) \ log \ \Sigma_{mol}}.
\end{equation}
This is consistent with the isolated spiral galaxies in both index and coefficient, as clearly seen in figure \ref{SchmidtMolAll}.
Interacting galaxies in the early stage obey the relation found in the isolated spiral galaxies.

Next, we look into a relationship between $\Sigma_\mathrm{SFR}$ and $\Sigma_\mathrm{gas}$ for our sample of interacting galaxies in the early and the mid stage.
Figure \ref{SchmidtGasAll} illustrates a power-law relation, which is seen as the case of $\Sigma_\mathrm{SFR}$ and $\Sigma_\mathrm{mol}$.
\begin{figure}[tbp]
	\centering
	\includegraphics{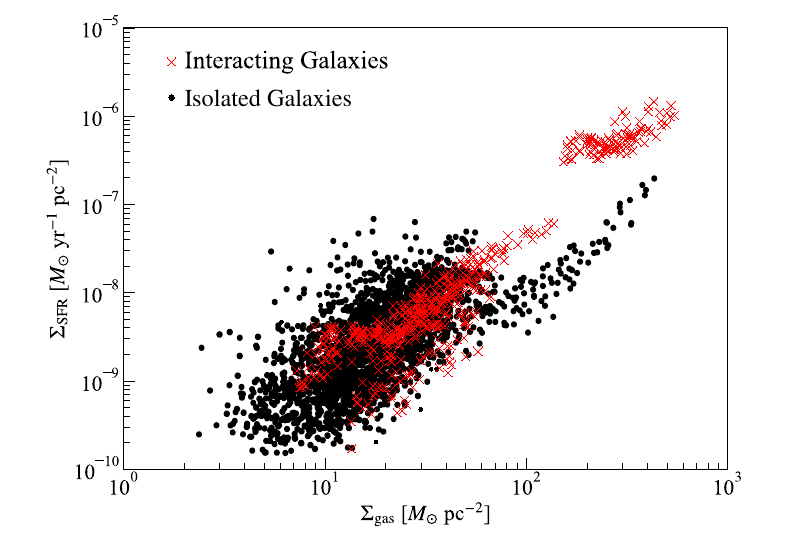}
	\caption{The same as figure \ref{SchmidtMolAll} but the relationship between $\Sigma_\mathrm{SFR}$ and $\Sigma_\mathrm{gas}$.} 
	\label{SchmidtGasAll}
\end{figure}
By using all pixels, interacting galaxies shows a power-law relation with an index of 1.84$\pm$0.03.

Figure \ref{SchmidtTot} represents the relationship between $\Sigma_\mathrm{SFR}$ and $\Sigma_\mathrm{gas}$ for interacting galaxies.
The power-law relation still holds in the case of $\Sigma_\mathrm{SFR}$-$\Sigma_\mathrm{gas}$.
\begin{figure}[tbp]
	\centering
	\includegraphics{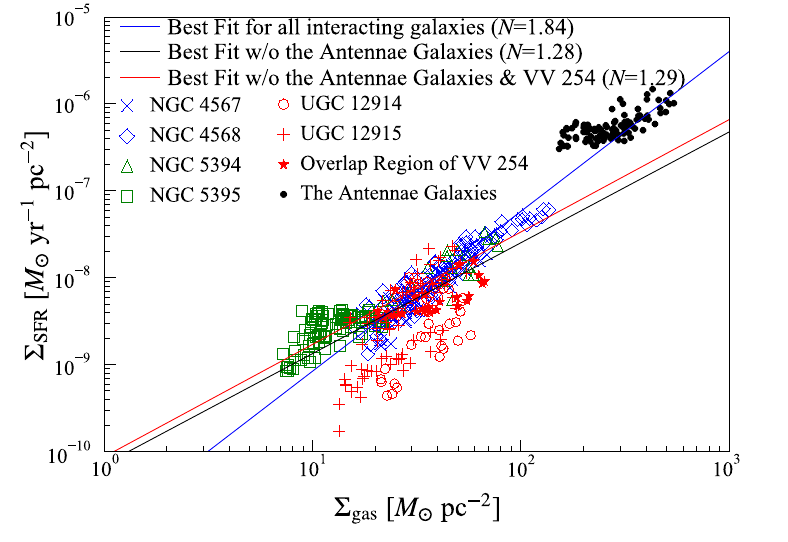}
	\caption{The same as figure \ref{SchmidtMol} but the relationship between $\Sigma_\mathrm{SFR}$ and $\Sigma_\mathrm{gas}$.}
	\label{SchmidtTot}
\end{figure}

As the inclination correction is not performed on $\Sigma_\mathrm{SFR}$ and $\Sigma_\mathrm{gas}$ for the Antennae Galaxies, we remove the data points of the Antennae Galaxies and fit again.
The interacting galaxies in the early stage of interaction have a power-law index of 1.28$\pm$0.06 with $R^{2}$ = 0.57.

VV\,254 have a large scatter, and some points in VV\,254 shows lower SFR (about 1 dex, which is significantly larger than the uncertainty) in $\Sigma_\mathrm{SFR}$-$\Sigma_\mathrm{mol}$ relation compared to the isolated galaxies (also see figure \ref{SchmidtGasAll}).
VV\,254 has the largest spatial resolution among the sample (5.8 kpc). 
As stated previously, a larger resolution makes a smaller scatter.
Therefore, a large scatter of VV\,254 cannot be explained by the difference of the angular resolution. 
This suggests that some physical factors cause a lower SFR in VV\,254.

Then, we exclude data points of VV\,254 in addition to that of the Antennae Galaxies.
This results in a tight power-law relation ($R^{2}$ = 0.85):
\begin{equation}
	{\rm log \ \Sigma_{SFR} = -(10.06\pm0.05) + (1.29\pm0.03) \ log \ \Sigma_{gas}}.
\end{equation}
The Kennicutt--Schmidt law is an ``averaged'' relationship between $\Sigma_\mathrm{SFR}$ and $\Sigma_\mathrm{ISM}$ over the focused region.
Therefore, the law is valid only when stars are formed constantly within the region.
In other words, the law becomes broken in a region where star formation activity is drastically changing in a short time (within a time scale that a used star-formation indicator traces).
The Antennae Galaxies and VV\,254 are good examples of this phenomenon.

To summarise our investigations about the Kenniccutt--Schmidt relation, we found a tight correlation with $\Sigma_\mathrm{SFR}$, whichever we use $\Sigma_\mathrm{mol}$ or $\Sigma_\mathrm{gas}$.
These Kennicutt--Schmidt laws are well consistent with isolated galaxies in the index and the coefficient, and they have super-linear indices ($N$ $>$ 1).
These facts imply that the Kennicutt--Schmidt law for the interacting galaxies in the early stage is not qualitatively changed compared with the isolated galaxies.
At least, the effect on star formation at the early stages of interaction is weaker than that of the bar structure.
However, we also note that the Antennae Galaxies, interacting galaxies in the mid stage, maybe in enhanced star-forming phase, and VV\,254, which is just after the high-velocity head-on collision, shows decreased star formation.

\subsection{Star formation history in interacting galaxies}
In this subsection, we discuss how the interaction affects ISM and star-forming activity based on the results of Paper II and this paper.
The main key findings are the high molecular gas fraction and local enhancement of SFEs in interacting galaxies even in the early stage, but no change of the index and the coefficients in the Kennicutt--Schmidt law.

When the galaxies start colliding, ISM environments are changed first.
During a collision of two galaxies, GMCs and H\emissiontype{I} clouds collide with each other and are then ionised, as discussed in Paper II.
Then, ionised gas with high pressure prevails over the discs and the overlap region along with the progress of the collision.
Radiative cooling makes ionised gas back into neutral hydrogen gas in a short time \citep{Harwit87}.
Such dense atomic hydrogen gas is converted into molecular hydrogen gas.
The conversion time scale depends on surrounding shocked gas pressure, determined by the initial collisional parameters.
As a result, the molecular gas fraction becomes higher during the collision because of the efficient transition from H\emissiontype{I} to H$_{2}$ gas.

However, at least in the early stages, there is no difference in the quality of star formation activity between interacting galaxies and isolated galaxies.
From the galaxy-scale Kennicutt--Schmidt law for the interacting galaxies and the isolated galaxies, the increase in SFR is found to be within a factor of three. 
Similarly, there is no difference in SFE on the galaxy scale.
The spatially-resolved Kennicutt--Schmidt law for interacting galaxies, which includes the data from severely disturbed galaxy NGC\,5395, has a tight correlation.
Its index and coefficient are the same as that for the isolated galaxies.

Star formation occurs from a diffuse molecular cloud via high-density molecular gas.
Observations of denser molecular gas tracers such as HCN and \atom{CO}{}{12}({\it J} = 3-2) revealed that SFE has a linear relation with a dense molecular gas fraction from giant molecular cloud to ULIRGs \citep{GS04,Michiyama16,Shimajiri17}.
Also, recent works of Our Galaxy showed that star formation activity has a better correlation with dense gas fraction than the amount of molecular gas \citep{Torii19}.
Therefore, even if many GMCs are formed through the conversion from H\emissiontype{I} to H$_{2}$ in interacting galaxies efficiently, star formation will not occur if there is not enough dense gas inside them.
Thus, high SFEs are achieved only when dense gas is efficiently produced on the target scale.
VV254 is low SFR and SFEs, although it has a large amount of molecular gas, which can be traced by CO.
The gas ionised by the collision returns to GMCs within a short time and becomes a large amount of molecular gas, but it is thought that it has not yet produced dense gas.
Low SFR and SFE in VV\,254 imply that VV\,254 are just ongoing the formation process of GMCs from ionised gas and not enough time for star formation yet.
When the time elapsed enough after the first collision, dense molecular gas could be formed at the colliding area (it could be the overlap region).
Local active star formation with high SFE occurs in that region, although the SFR and SFE in a galaxy scale are still low.

On the other hand, the situation changes in the later stage of interaction. 
As suggested from Paper II in the Antennae Galaxies, which are the most progressed interacting galaxies among our sample, the molecular gas fraction will get much higher throughout the pair.
At that phase, we may also expect that molecular gas becomes denser due to high gas concentration toward the centre (e.g., \cite{Scoville91}).
Dense gas in the galaxy centre leads to start starburst such as ULIRGs.
The galaxy-wide starburst would make galaxy-scale SFR and SFE high.

\section{Summary}
We investigated the relationship between ISM and star formation activity in interacting galaxies in the early and mid stages of the interaction using CO, H\emissiontype{I}, and dust-corrected star formation rate images. 
Our findings are: 
\begin{enumerate}
	\item{On a galactic scale, interacting galaxies have a similar sSFR compared with the isolated sample.}
	\item{The galactic-scale Kennicutt--Schmidt law reveals that most of our samples do not show enhanced SFR.}
	\item{Galaxy-scale sSFR and SFE of the interacting galaxies are comparable to those of the isolated spiral galaxies.}
	\item{SFE maps show asymmetric distributions or local peaks at off-centre regions.}
	\item{Higher SFE in smaller progenitors could be due to largely-prevailed shock.}
	\item{Difference of SFE at the collision front can be explained by the time after the collision and strength of shock depending on collision parameters.}
	\item{No difference of the index and coefficient in kpc-scale Kennicutt--Schmidt law in the early stage is found between interacting galaxies in the early stage and isolated galaxies, although interacting galaxies have higher molecular gas fraction than isolated galaxies.}
\end{enumerate}

\bigskip
\section*{Funding}
This work was supported by Japan Society for the Promotion of Science KAKENHI (Grant No. 18K13593).

\section*{Acknowledgement}
We would like to thank all staff members of NRO for observational support. This research made use of the NASA/IPAC
Extragalactic Database (NED), which is operated by the Jet Propulsion Laboratory, California Institute of Technology, under contract with the National Aeronautics and Space Administration.

\appendix
\section*{Effect of Oversampling on the Kennicutt--Schmidt law}\label{appendices}
Oversampling may affect the results of analyses.
We check the robustness of our results by using two dummy maps that reproduce the index, coefficient, and scatter of the observed Kennicutt--Schmidt law.
In this analysis, the first map, ``map $\alpha$'', mimics the surface density of ISM ($\Sigma_\mathrm{gas}$), while the second map, ``map $\beta$'', imitates the surface density of SFR ($\Sigma_\mathrm{SFR}$).
The two-dimensional dummy maps are made as the following steps.
\begin{enumerate}
\item An original source map is made to have a size of 21 $\times$ 21 pixels. 
	We embed ``sources'', which have $\Sigma_\alpha$, a value of the map $\alpha$, at $i$-th column and $j$-th row
	\begin{equation}
		\Sigma_\alpha(i, j) = \left\{
		\begin{array}{ll}
			10^{1.3-0.2[abs(11-i)]}\times10^{1.3-0.2[abs(11-j)]} & (6\le i \le16, 6\le j \le16)\\
			0 & (\mbox{otherwise})
		\end{array}
		\right..
	\end{equation}
These values are set to have the similar dynamic range of the observed $\Sigma_\mathrm{gas}$ (1.0 $\lesssim \log \Sigma_\mathrm{gas} \lesssim$ 2.5).
\item The map $\beta$ is made to obey a power-law relation as follows:
\begin{equation}
	\rm log \ \Sigma_\beta = -10.0 + 1.3 \ \log \ \Sigma_\alpha 
	\label{mimic_equation}
\end{equation}
\item We add Gaussian-like noises onto ``source'' maps to represent thermal noise.
Noises are set to have as same as the observed $\Sigma_\mathrm{gas}$ and $\Sigma_\mathrm{SFR}$.
\item The map $\alpha$ and $\beta$ are convolved with a Gaussian, which has FWHM of 1 pixel, to represent observations with a Gaussian-shaped beam.
\item In order to replicate data reduction due to On-The-Fly technique, the map $\alpha$ is convolved with a Bessel--Gaussian kernel.
The Bessel-Gaussian kernel has implemented to have 
\begin{equation}
	f(r) = \left\{
	\begin{array}{ll}
		\frac{J_1(\pi r/a)}{\pi r/a}
		\exp \left[ -\left( \frac{r}{b} \right)^2 \right] &
		(r \le R_{\rm max}) \\
		0 & (r > R_{\rm max}).
	\end{array}
	\right.
\end{equation}
where $J_{1}$ is a first-order Bessel function, $r$ is the distance between the data point in the unit of the grid spacing.
The parameters $a$, $b$, and $R_\mathrm{max}$ are set as $a$ = 1.55, $b$ = 2.52, and $R_\mathrm{max}$ = 3, which are as same as the parameters adopted in the data reduction of the CO data.
With these parameters, the convolution yields an effective beam, which has 1.29 times larger beam size than the beam made in the step 4.
\end{enumerate}
We make a hundred sets of these maps and derive the correlation coefficient $R^{2}$ between the map $\alpha$ and map $\beta$ with a Nyquist rate (a sampling rate of every 2 pixels) and oversampling (a sampling rate of every 2.58 pixels).
We mask the emission lower than 3 $\sigma$ for deriving $R^{2}$.
One of $\Sigma_\alpha$-$\Sigma_\beta$ plots with the Nyquist sampling and oversampling is shown in figure \ref{example_alpha_beta}.
\begin{figure}[tbp]
	\centering
	\includegraphics{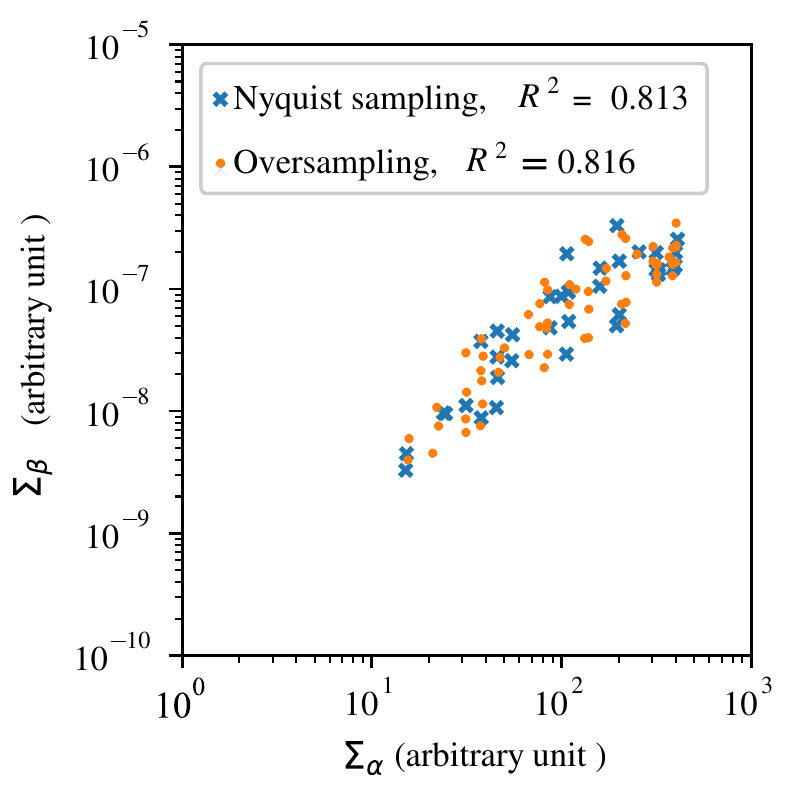}
	\caption{An example of relationship between $\Sigma_\alpha$ and $\Sigma_\beta$ with the Nyquist sampling and oversampling.}
	\label{example_alpha_beta}
\end{figure}
The plot illustrates the dummy maps successfully reproduce the observed Kennicutt--Schmidt law.
Although $R^{2}$ is not a ratio scale but an ordinal scale, checking $R^{2}$ distributions is effective to see the effect of difference of sampling.
We derived $\Delta R^2\equiv R^2_\mathrm{oversampling}-R^2_\mathrm{Nyquist}$, where $R^2_\mathrm{oversampling}$ is $R^2$ for the oversampling case and $R^2_\mathrm{Nyquist}$ is $R^2$ for the Nyquist sampling case.
\begin{figure}[tbp]
	\centering
	\includegraphics{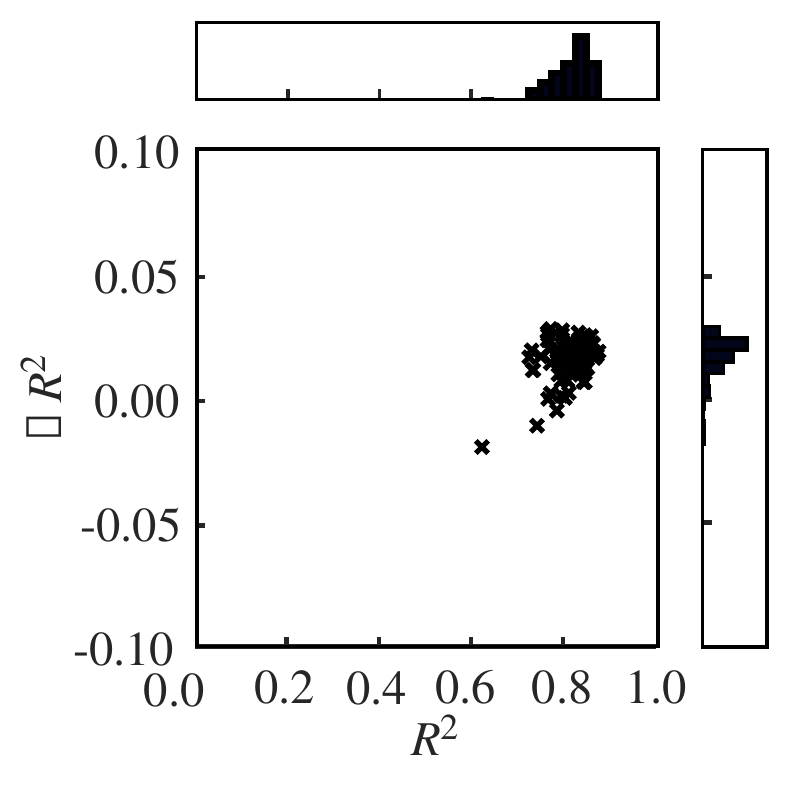}
	\caption{Scatter plot of $R^2$-$\Delta R^2$ and their histograms.}
	\label{R2-dR2}
\end{figure}
\begin{figure}[tbp]
	\centering
	\includegraphics{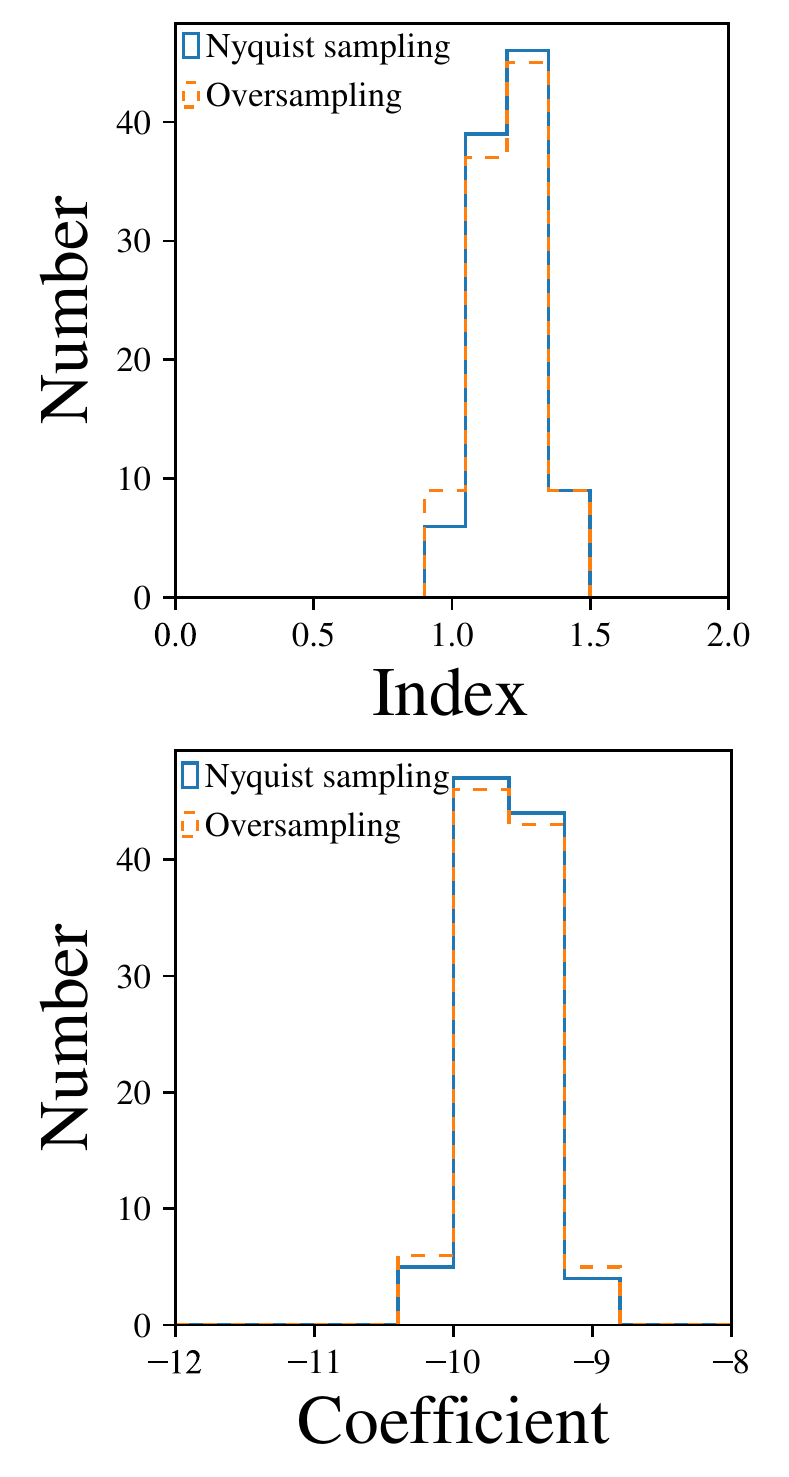}
	\caption{Histograms of index and coefficient of the fits for mock images.}
	\label{hist_indices_coefficients}
\end{figure}
As shown in figure \ref{R2-dR2}, the dummy maps correlate with $R^2\sim$ 0.8, and $\Delta R^2$ show an average of 0.02 in those cases.
We also investigate the effect of the difference of sampling rates for indices and coefficients.
Figure \ref{hist_indices_coefficients} shows that the indices and coefficients are not changed for difference of samplings.
These findings mean if we sample a map with oversampling, the fitting result is slightly better than the case that the data is sampled with the Nyquist sampling in most cases ($\sim$90 cases).
However, the sampling effect does not change the significance of the result.
Therefore, our findings about the Kennicutt--Schmidt law should be real.

\end{document}